\documentclass[12pt]{article} 
\usepackage[a4paper,textwidth=490.0pt,textheight=703.1pt]{geometry}
\usepackage{url}
\usepackage{booktabs}

\usepackage{epsfig} %
\usepackage{float} %
\usepackage{amssymb} %
\usepackage{amsmath} %
\usepackage{verbatim} %
\usepackage{cite} %
\usepackage{color} %
\usepackage{eufrak}
\usepackage[english]{babel}
\usepackage[latin1]{inputenc}
\usepackage[T1]{fontenc}
\usepackage{ae}
\usepackage{url}   
\usepackage{graphicx} 
\usepackage{booktabs}
\usepackage{slashed}

\newcommand{\HA}{{\rm H}}

\newcommand{\ep}{\varepsilon}

\newcommand{\beq}{\begin{equation}}
\newcommand{\eeq}{\end{equation}}
\newcommand{\bea}{\begin{eqnarray}}
\newcommand{\eea}{\end{eqnarray}}

\allowdisplaybreaks[4]

\begin{document} 
\setlength{\baselineskip}{0.515cm}

\sloppy 
\thispagestyle{empty} 
\begin{flushleft} 
DESY 21--104
\\ 
DO--TH 21/23
\\ 
TTP 21--024
\\ 
RISC Report Series 21--13
\\
SAGEX--21--15
\\ 
July 2021 
\end{flushleft}

\mbox{} \vspace*{\fill} \begin{center}

{\Large\bf The three-loop unpolarized and polarized}

\vspace*{2mm} 
{\Large\bf non-singlet anomalous dimensions from}

\vspace*{2mm} 
{\Large\bf off shell operator matrix elements}

\vspace{3cm} 
\large 
{\large J.~Bl\"umlein$^a$, P.~Marquard$^a$, C.~Schneider$^b$ and K.~Sch\"onwald$^{c}$ }

\normalsize 

\vspace{1.cm} 
{\it $^a$~Deutsches Elektronen--Synchrotron DESY,}\\ {\it Platanenallee 6, D--15738 Zeuthen, Germany}

\vspace*{2mm} 
{\it $^b$~
Johannes Kepler University Linz, Research Institute for Symbolic
Computation (RISC) ,
Altenberger Stra{\ss}e 69, A-4040 Linz, Austria}

\vspace*{2mm} 
{\it $^c$~Institut f\"ur Theoretische Teilchenphysik,\\ Karlsruher Institut f\"ur Technologie (KIT) D--76128 
Karlsruhe, Germany}


\end{center} 
\normalsize 
\vspace{\fill} 
\begin{abstract} 
\noindent
We calculate the unpolarized and polarized three--loop anomalous dimensions and splitting functions 
$P_{\rm NS}^+, P_{\rm NS}^-$ and $P_{\rm NS}^{\rm s}$ in QCD in the $\overline{\sf MS}$ scheme by using 
the traditional method of space--like off shell massless operator matrix elements. This is a gauge--dependent 
framework. For the first time we also calculate the three--loop anomalous dimensions $P_{\rm NS}^{\rm \pm, tr}$ 
for transversity directly. We compare our results to the literature. 
\end{abstract}

\vspace*{\fill} \noindent
\newpage

\section{Introduction} 
\label{sec:1}

\vspace*{1mm} 
\noindent 
The anomalous dimensions of local quark and gluon operators determine the scaling violations of 
the deep--inelastic scattering structure functions \cite{Politzer:1974fr,REV} by the scale evolution of the 
parton densities and are therefore instrumental in the measurement of the strong coupling constant 
$a(M_Z^2) = \alpha_s(M_Z^2)/(4\pi)$ \cite{alphas} for this inclusive precision data. They have been calculated 
to 3--loop order both in the unpolarized and polarized case \cite{Moch:2004pa,Vogt:2004mw,Moch:2014sna,Moch:2015usa} 
using the method of on--shell forward Compton amplitudes, in which the scale is set by the virtuality $Q^2 = -q^2$ 
of the exchanged current. At four--loop order a series of low moments for the non--singlet anomalous dimensions has 
been calculated in Refs.~\cite{FOUR} and at five--loop order in \cite{Herzog:2018kwj}.
The $O(T_F)$ contributions at three--loop order have been confirmed by the 
calculation of massive on-shell operator matrix elements (OMEs) \cite{Ablinger:2014vwa,Ablinger:2014lka,
Ablinger:2014nga,Ablinger:2017tan,Behring:2019tus}.

The traditional way of calculating the anomalous dimensions consists in computing the off shell
massless local OMEs, cf.~\cite{Gross:1973ju,Gross:1974cs,Georgi:1951sr,Sasaki:1975hk,Ahmed:1975tj}
in the one--loop case, which in general implies the breaking of gauge invariance to be dealt with.
The two--loop anomalous dimensions have been calculated in 
\cite{Floratos:1977au,GonzalezArroyo:1979he,GonzalezArroyo:1979ng,GonzalezArroyo:1979df,
Curci:1980uw,Furmanski:1980cm,Floratos:1981hs,Hamberg:1991qt,HAMBERG,Mertig:1995ny,SP_PS1,Ellis:1996nn,
Matiounine:1998ky,Matiounine:1998re,Moch:1999eb,Vogt:2008yw,Ablinger:2014vwa,Ablinger:2014lka,
Ablinger:2014nga,Ablinger:2017tan,Behring:2019tus}.

In this paper we are calculating the unpolarized and polarized three--loop anomalous 
dimensions for the first 
time using the method of massless off shell OMEs in the flavor non--singlet case, 
which is the first complete independent recalculation of the results obtained in 
Ref.~\cite{Moch:2004pa}. The present calculation requires the 
knowledge of the corresponding massless off shell OMEs to two--loop order, 
cf.~\cite{Matiounine:1998ky,Matiounine:1998re,TWOLOOP}, up to the terms of $O(\ep^0)$ in the dimensional 
parameter $\ep =D-4$. The off shell OMEs are gauge--dependent quantities. We will calculate the 
anomalous dimensions and splitting functions: $P_{\rm NS}^+, P_{\rm NS}^-$ and $P_{\rm NS}^{\rm s}$. 
For the first time we also calculate the three--loop anomalous dimension $P_{\rm NS}^{\rm \pm, tr}$ for transversity 
in a direct way.

The paper is organized as follows. In Section~\ref{sec:2} we derive the structure of the physical part
of the flavor non--singlet unrenormalized off shell OMEs to three--loop order. From their pole terms of 
$O(1/\ep)$ one can extract the non--singlet anomalous dimensions. Due to a known Ward identity, cf. e.g.
\cite{Ablinger:2014vwa,Behring:2019tus}, the polarized anomalous dimension can be calculated by applying 
anticommuting $\gamma_5$. We also calculate the polarized OMEs in the Larin scheme \cite{Larin:1993tq,Matiounine:1998re}
from which one can determine the $Z$--factor $Z_5^{\rm NS}(N)$ of the corresponding finite renormalization to three--loop 
order. The details of the calculation are described in Section~\ref{sec:3}. In Section~\ref{sec:4} we 
present the three--loop anomalous dimensions and splitting functions. We compare with results in the 
literature in Section~\ref{sec:5} and Section~\ref{sec:6} contains the conclusions. In an appendix we 
briefly summarize the transition from the Larin to the $\overline{\sf MS}$ scheme for the polarized anomalous 
dimension in the vector case.

\section{The unrenormalized operator matrix elements} 
\label{sec:2}

\vspace*{1mm} 
\noindent 
The massless off shell non--singlet OMEs are defined as expectation values of the local operators
\begin{eqnarray}
\label{eq:op1}
O_{q,r; \mu_1 ... \mu_N}^{\rm NS} &=&
i^{N-1} {\rm\bf S}\Biggl[
\overline{\psi} \gamma_{\mu_1} D_{\mu_2} ... D_{\mu_N} \frac{\lambda_r}{2} \psi \Biggr] -~\text{trace~terms},
\\
\label{eq:op2}
O_{q,r; \mu_1 ... \mu_N}^{\rm NS,5} &=&
i^{N-1} {\rm\bf S}\Biggl[
\overline{\psi} \gamma_5  \gamma_{\mu_1} D_{\mu_2} ... D_{\mu_N} \frac{\lambda_r}{2} \psi \Biggr] -~\text{trace~terms}
\end{eqnarray}
between quark (antiquark) states $\psi~(\bar{\psi})$ of space--like momentum $p$, $p^2 < 0$, and are given by 
\begin{eqnarray}
\hat{A}_{qq}^{\rm NS, (5)} = \langle q(p)|O^{\rm NS, (5)}|q(p)\rangle.
\end{eqnarray}
Here {\bf S} is the symmetry operator, $\lambda_r$ a $SU(N_F)$ flavor matrix and $D_\mu = \partial_\mu 
+ i g_s t_a A_\mu^a$ the covariant derivative, with $ A_\mu^a$ the gluon field,  $\psi$  the 
quark field, $t_a$ the generators of $SU(N_C)$, and $g_s = \sqrt{4 \pi \alpha_s}$. The Feynman rules of QCD are 
given in \cite{YND} and for the local operators in 
\cite{Bierenbaum:2009mv,Behring:2019tus}.
The local operator in the case of transversity is given by
\begin{eqnarray}
\label{eq:op3}
O_{q,r; \mu \mu_1 ... \mu_N}^{\rm NS, tr} &=&
i^{N-1} {\rm\bf S}\Biggl[
\overline{\psi} \sigma_{\mu \mu_1} D_{\mu_2} ... D_{\mu_N} \frac{\lambda_r}{2} \psi \Biggr] 
-~\text{trace~terms}~,
\end{eqnarray}
where $\sigma_{\mu \nu} = (i/2)[\gamma_\mu \gamma_\nu - \gamma_\nu \gamma_\mu]$.

The operator matrix elements have the representation
\begin{eqnarray}
\label{eq:Aiq}
\hat{A}_{qq}^{\rm NS} = \left[ \Delta \hspace*{-2.5mm} \slash \hat{A}_{qq}^{\rm NS, phys} + p 
\hspace*{-2mm} \slash \frac{\Delta.p}{p^2} 
\hat{A}_{qq}^{\rm NS, EOM} \right] (\Delta.p)^{N-1}.
\end{eqnarray}
Here $\Delta$ denotes a light--like vector, $\Delta.\Delta = 0$. 
The following projectors are applied to separate the physical (phys) contribution and the one 
vanishing by the equation of motion (EOM), which does not hold in the off shell case,
\begin{eqnarray}
\hat{A}_{qq}^{\rm NS, phys} &=& \frac{1}{4 (\Delta.p)^{N}}~\text{tr}~\Biggl[\Biggl(p \hspace*{-2mm} 
\slash - 
\frac{p^2}{\Delta.p} 
\Delta 
\hspace*{-2.5mm} \slash\Biggr) \hat{A}_{qq}^{\rm NS} \Biggr],
\\
\hat{A}_{qq}^{\rm NS, EOM} &=& \frac{1}{4 (\Delta.p)^{N}}~\text{tr}~\Biggl[\Delta \hspace*{-2.5mm} 
\slash  
\hat{A}_{qq}^{\rm NS} \Biggr].
\end{eqnarray}
In the polarized case the operator (\ref{eq:op1}) is replaced by the operator of 
Eq.~(\ref{eq:op2}) and the following relations hold 
\begin{eqnarray}
    \hat{A}_{qq}^{\rm NS,5} = \left[ \gamma_5 \Delta \hspace*{-2.5mm} \slash \hat{A}_{qq}^{\rm NS,5, phys}
    + \gamma_5 p \hspace*{-2mm} \slash \frac{\Delta.p}{p^2}
    \hat{A}_{qq}^{\rm NS,5, EOM} \right] (\Delta.p)^{N-1}
\end{eqnarray}
with
\begin{eqnarray}
    \hat{A}_{qq}^{\rm NS,5,phys} &=& \frac{1}{4 (\Delta.p)^{N}}~\text{tr}~\Biggl[\Biggl(p \hspace*{-2mm}
    \slash - \frac{p^2}{\Delta.p}
    \Delta \hspace*{-2.5mm} \slash\Biggr)
    \gamma_5 \hat{A}_{qq}^{\rm NS,5} \Biggr],
\\
    \hat{A}_{qq}^{\rm NS,5,EOM} &=&
    \frac{1}{4 (\Delta.p)^{N}}~\text{tr}~\Biggl[\Delta \hspace*{-2.5mm}
    \slash  \gamma_5 \hat{A}_{qq}^{\rm NS,5} \Biggr].
\end{eqnarray}

In the case of transversity we consider the unrenormalized Green's function \cite{Blumlein:2009rg}
\begin{eqnarray}
\hat{G}_{\mu,qQ}^{ij,\rm NS,tr} &=& \delta_{ij} (\Delta.p)^{N-1} \Biggl[
\Delta_\rho \sigma^{\mu \rho} \Delta_T \hat{A}_{qq}^{\rm NS,phys}\left(\frac{-p^2}{\mu^2}, \ep, N \right)
+ c_1 \Delta^\mu + c_2 p^\mu + c_3 \gamma_\mu p \hspace*{-1.8mm} \slash + c_4 \Delta \hspace*{-2.2mm} 
\slash p \hspace*{-1.8mm} \slash \Delta^\mu
\nonumber\\ &&
+ c_5 \Delta \hspace*{-2.2mm} \slash p \hspace*{-1.8mm} \slash p^\mu 
\Biggr],
\end{eqnarray}
where $i,j$ are external color indices  and the coefficients $\left. c_k\right|_{k= 1...5}$ denote other OMEs than 
those we are going to deal with.
 
Since the non--singlet anomalous dimensions receive only contributions from the unrenormalized OME 
$\hat{A}_{qq}^{\rm NS,(5), phys}$ we will consider only this operator matrix element in the 
following. In Mellin $N$ space it has the representation 
\begin{eqnarray}
\hat{A}_{qq}^{\rm NS, (5)} &=& 1 + \sum_{k=1}^\infty \hat{a}^k 
S_\ep^k \left(\frac{-p^2}{\mu^2}\right)^{\ep k/2} \hat{A}_{qq}^{(k), \rm NS, (5)},
\end{eqnarray}
with the spherical factor
\begin{eqnarray}
S_\ep = \exp\left[\frac{\ep}{2}(\gamma_E - \ln(4\pi))\right],
\end{eqnarray}
where $\gamma_E$ is the Euler--Mascheroni number  and $\hat{a}$  the bare coupling 
constant.
The free gluon propagator is given by\footnote{Note a typo in \cite{Matiounine:1998ky}, 
Eq.~(2.6).}
\begin{eqnarray}
D_{\mu\nu}^{ab}(k) = \frac{\delta^{ab}}{k^2 + i0} \left[-g_{\mu\nu} + (1- \hat{\xi}) 
\frac{k_\mu k_\nu}{k^2 + 
i0}\right],
\end{eqnarray}
which defines the gauge parameter in the $R_{\hat{\xi}}$ gauge. The renormalization of the massive off shell 
non--singlet OMEs encounters the renormalization of the coupling constant and the gauge parameter, as well as 
that of the local operator. In the following we will deviate from Refs.~\cite{Matiounine:1998ky,Matiounine:1998re} and 
perform the renormalization of the coupling constant and the gauge parameter and use the 
resulting expression, $\tilde{A}$,
at $\mu^2 = - p^2$ to extract the anomalous dimensions. In the unrenormalized OME 
obtained in the diagrammatic
calculation the coupling constant and the gauge parameter are renormalized before comparing to $\tilde{A}$ in 
Eq.~(\ref{eq:Ahat3un}).
The unrenormalized coupling is given by
\begin{eqnarray}
\hat{a} = a \left[1 + \frac{2}{\ep} \beta_0 a + \left(\frac{4}{\ep^2} \beta_0^2 + \frac{1}{\ep} \beta_1\right) a^2\right]
+ O(a^3),
\end{eqnarray}
where $a$ denotes the renormalized strong coupling constant.
The expansion coefficients of the QCD $\beta$--function are given by \cite{BETA}\footnote{Note 
a typographical error in \cite{Matiounine:1998ky}, Eq.~(2.13) and \cite{Matiounine:1998re}, Eq.~(2.14).}
\begin{eqnarray}
\beta_0 &=& \frac{11}{3} \textcolor{blue}{C_A} - \frac{4}{3} \textcolor{blue}{T_F N_F},
\\
\beta_1 &=& \frac{34}{3} \textcolor{blue}{C_A^2} - \frac{20}{3} 
\textcolor{blue}{C_A T_F N_F} - 4 \textcolor{blue}{C_F T_F N_F}.
\end{eqnarray}
The bare gauge parameter $\hat{\xi}$ is renormalized by
\begin{eqnarray}
  \hat{\xi} = \xi~Z_3(\xi),
\end{eqnarray}
where $Z_3$ is the $Z$--factor of the gluon propagator,
cf. \cite{Egorian:1978zx,Chetyrkin:2004mf,Luthe:2016xec,
Chetyrkin:2017bjc},
\begin{eqnarray}
Z_3(\xi) &=& 1 + a \frac{z_{11}}{\ep} + a^2 \Biggl[ \frac{z_{22}}{\ep^2} +\frac{z_{21}}{\ep}\Biggr] + 
O(a^3),
\end{eqnarray}
with
\begin{eqnarray}
z_{11} &=& \textcolor{blue}{C_A} \left[-\frac{13}{3} + {\xi}\right] + \frac{8}{3} 
\textcolor{blue}{T_F N_F},
\\
z_{22} &=& \textcolor{blue}{C_A^2} \left[-\frac{13}{2} - \frac{17}{6} {\xi} + {\xi}^2\right] + 
\textcolor{blue}{C_A T_F N_F}\left[4 + \frac{8}{3} {\xi}\right],  
\\
z_{21} &=& \textcolor{blue}{C_A^2} \left[-\frac{59}{8} + \frac{11}{8} {\xi} + \frac{1}{4} 
{\xi}^2 
\right]
+ 4~\textcolor{blue}{C_F T_F N_F} 
+ 5~\textcolor{blue}{C_A T_F N_F}. 
\end{eqnarray}
The color factors are $\textcolor{blue}{C_F} = (N_C^2-1)/(2 N_C), \textcolor{blue}{C_A} = N_C, 
\textcolor{blue}{T_F} = 1/2$ for $SU(N_C)$ and $N_C = 3$ for QCD; $\textcolor{blue}{N_F}$ denotes the 
number of massless quark flavors.

In Mellin $N$ space the $Z$-factor of a local non--singlet operator reads \cite{Bierenbaum:2009mv} 
\begin{eqnarray}
Z^{\rm NS} &=& 1 
               + a \frac{\gamma_{qq}^{(0),\rm NS}}{\ep}
               + a^2 \Biggl[\frac{1}{\ep^2} \Biggl(\frac{1}{2} {\gamma_{qq}^{(0),\rm NS}}^2 +
                     \beta_0 \gamma_{qq}^{(0),\rm NS}\Biggr) + \frac{1}{2\ep} \gamma_{qq}^{(0),\rm NS} \Biggr]
+ a^3 \Biggl[ \frac{1}{\ep^3} \Biggl( \frac{1}{6} {\gamma_{qq}^{(0),\rm NS}}^3 
\nonumber\\ &&
+ \beta_0 {\gamma_{qq}^{(0),\rm NS}}^2 + \frac{4}{3} \beta_0^2 \gamma_{qq}^{(0),\rm NS}\Biggr)
+ \frac{1}{\ep^2} \Biggl( \frac{1}{2} \gamma_{qq}^{(0),\rm NS} \gamma_{qq}^{(1),\rm NS}
+ \frac{2}{3} \beta_0 \gamma_{qq}^{(1),\rm NS} + \frac{2}{3} \beta_1 \gamma_{qq}^{(0),\rm NS} \Biggr) 
\nonumber\\ &&
+ \frac{1}{3 \ep} \gamma_{qq}^{(2),\rm NS}\Biggr].
\label{eq:ZNS}
\end{eqnarray}
In (\ref{eq:ZNS}) the terms $\gamma_{qq}^{(k),\rm NS},~~k = 0,1,2, \ldots$ denote the expansion 
coefficients of the
anomalous dimension  
\begin{eqnarray}
\gamma_{\rm NS} = \sum_{k=1}^\infty a^k \gamma_{\rm NS}^{(k-1)}.
\end{eqnarray}
The partly renormalized OME, $\tilde{A}_{qq}^{\rm NS, phys}$, reads
\begin{eqnarray}
\tilde{A}_{qq}^{\rm NS, phys} &=& 1 + a \Biggl[
  \frac{a_{qq}^{\rm NS,(1,-1)}}{\ep}
+ a_{qq}^{\rm NS,(1,0)}
+ a_{qq}^{\rm NS,(1,1)} \ep \Biggr]
+ a^2 \Biggl[
  \frac{a_{qq}^{\rm NS,(2,-2)}}{\ep^2}
+ \frac{a_{qq}^{\rm NS,(2,-1)}}{\ep}
+ a_{qq}^{\rm NS,(2,0)}  \Biggr]
\nonumber\\ &&
+ a^3 \Biggl[
  \frac{a_{qq}^{\rm NS,(3,-3)}}{\ep^3}
+ \frac{a_{qq}^{\rm NS,(3,-2)}}{\ep^2}
+ \frac{a_{qq}^{\rm NS,(3,-1)}}{\ep}
  \Biggr].
\label{eq:Ahat3un}
\end{eqnarray}
The expansion coefficients $a_{qq}^{{\rm NS}, (i,j)}$ are in general gauge dependent. 
The renormalized OMEs are given by
\begin{eqnarray}
{A}_{qq}^{\rm NS, phys} &=& \frac{\tilde{A}_{qq}^{\rm NS, phys}}{Z^{\rm NS}},
\end{eqnarray}
expanded to $O(a^3)$ and setting $S_\ep = 1$. The anomalous dimensions are iteratively extracted form the $1/\ep$ 
pole terms and the 
other expansion coefficients $a_{qq}^{{\rm NS}, (i,j)}$ are given in Ref.~\cite{TWOLOOP}.

Eq.~(\ref{eq:Ahat3un}) is understood to hold both for the unpolarized as well as the polarized case, by relabeling
the corresponding quantities to $f \rightarrow \Delta f$. Similar expressions 
hold for transversity. From them we will determine $\gamma_{\rm NS}^{(2)}$ and $\Delta \gamma_{\rm NS}^{(2)}$
in both cases. The further
three--loop non--singlet anomalous dimensions $\gamma_{\rm NS}^{(2),\rm s}$ can be derived from other quarkonic 
diagrams at three--loop order.\footnote{There is a further non--singlet anomalous dimension
$\Delta \gamma_{\rm NS}^{(2),\rm s}$ \cite{Moch:2015usa} occurring in the pole--term of an axialvector--vector 
current interference 
contribution in the forward Compton amplitude, related to the polarized structure function $g_5$, introduced in 
Ref.~\cite{Blumlein:1996vs}, which has {\it even} moments. These aspects are of  importance but are not discussed in 
\cite{PDG}. We will consider this quantity elsewhere.} 
Because $\gamma_{\rm NS}^{(2),\rm s}$ occurs for the first time at the three--loop loop level,
there is no renormalization of the OME
\begin{eqnarray}
\frac{2}{1 + (-1)^N} \left. \hat{A}_{qq}^{\rm PS,(3), phys}(N) \right|_{d_{abc} d^{abc}}  = \hat{a}_s^3 
\frac{1}{3 \ep} 
\gamma_{\rm NS}^{\rm (2),s}(N) + O(\ep^0),~~N \in \mathbb{N},~{\rm odd}, N \geq 1. 
\end{eqnarray}
The other expansion coefficients occurring in (\ref{eq:Ahat3un}) potentially
coming from lower orders in the coupling $a$
do all vanish in this case. The anomalous dimension $\gamma_{\rm NS}^{(2) \rm s}$ 
is formally obtained as the $O(1/\ep)$ pole term of the 
pure--singlet OME by considering in the unpolarized case the analytic continuation from odd 
values of $N$. The $d_{abc} d^{abc}$ terms in the non--singlet $+$ contributions vanish. 
One considers 
the contributions $\propto d_{abc} d^{abc}$ of this OME for the {\it odd} moments.
In this way $\gamma_{\rm NS}^{\rm (2),s}$ corresponds to the non--singlet
combination $\gamma_{\rm qq}^{\rm (2),s} - 
\gamma_{\rm q\bar{q}}^{\rm (2),s}$.

In deep--inelastic scattering one may form up to three different combinations of quark distributions in the
unpolarized and polarized case
\begin{eqnarray}
q_{{\rm NS},ik}^+ &=& q_i + \bar{q}_i - (q_k + \bar{q}_k),
\\
q_{{\rm NS},ik}^- &=& q_i - \bar{q}_i - (q_k - \bar{q}_k),
\\
\label{eq:PAR3}
q^{\rm v}_{\rm NS} &=& \sum_{k=1}^{N_F} (q_k - \bar{q}_k)
\end{eqnarray}
and analogously for $(q_k, \bar{q}_k) \rightarrow (\Delta q_k, \Delta \bar{q}_k)$. Here $i,k$ 
denote the different flavors. These combinations can be obtained by combining the scattering cross 
sections for different neutral and charged current exchanges off proton and neutron 
targets.\footnote{In the case of deuteron or He$^3$ targets nuclear wave function corrections have to 
be applied. Heavier nuclear targets have quite a variety of different corrections, known as EMC 
effect \cite{Rith:1983tw}.}
The corresponding anomalous dimensions ruling the evolution of these non--singlet distributions are
$\gamma_{\rm NS}^+, \gamma_{\rm NS}^-$ and $\gamma_{\rm NS}^{\rm v} = \gamma_{\rm NS}^- + \gamma_{\rm 
NS}^{\rm s}$. In the polarized case
mostly pure virtual photon exchange has been studied experimentally, which is described by  the structure 
functions $g_{1,2}(x,Q^2)$. Their non--singlet contributions evolve with $\Delta \gamma_{\rm NS}^+$. 
The following relations hold 
\begin{eqnarray}
\Delta \gamma_{\rm NS}^+ &=&  \gamma_{\rm NS}^-,
\\
\Delta \gamma_{\rm NS}^- &=&  \gamma_{\rm NS}^+.
\end{eqnarray}
\section{Details of the calculation}
\label{sec:3}

\vspace*{1mm} 
\noindent 
The Feynman diagrams for the massless off shell OMEs are generated by {\tt QGRAF} 
\cite{Nogueira:1991ex,Bierenbaum:2009mv} and the Dirac and Lorentz algebra is performed by 
{\tt FORM} \cite{FORM}.
The color algebra is performed by using {\tt Color} \cite{vanRitbergen:1998pn}.
The local operators are resummed into propagators by observing the current crossing relations, 
cf.~\cite{Politzer:1974fr,Blumlein:1996vs}, as has been described in Ref.~\cite{TWOLOOP}, in the 
corresponding OMEs $\hat{A}_{qq}^{{\rm NS},(5)}$ for even or odd moments, which will depend on the 
resummation variable $t$ quadratically only. To calculate the anomalous dimension 
$\gamma_{\rm NS}^{(2),s}$ we resum first, using the variable $t$ itself.

In the flavor non--singlet case 684 irreducible diagrams contribute. The reducible diagrams are accounted 
for by wave--function renormalization \cite{Chetyrkin:2017bjc,Egorian:1978zx,Luthe:2016xec}, decorating the OMEs at 
lower order in the coupling constant \cite{Matiounine:1998ky,Matiounine:1998re,TWOLOOP}. The different local 
operator insertions are resummed using generating functions of the type
\begin{eqnarray}
\sum_{N=0}^\infty (\Delta.k)^N \left(t^N \pm (-t)^N\right) \rightarrow
\left[\frac{1}{1 - \Delta.k~t} \pm  \frac{1}{1 + \Delta.k~t}\right],
\label{eq:resu1}
\end{eqnarray}
where $t$ denotes an auxiliary parameter for the resummation of the formal Taylor series, see \cite{Ablinger:2014yaa}.
Eq.~(\ref{eq:resu1}) implements the corresponding current crossing relations in the unpolarized (+) and the
polarized case $(-)$ \cite{Politzer:1974fr,Blumlein:1996vs}, which is not just a formality. Only the moments contributing 
to the respective cases exist.\footnote{This representation is sometimes misinterpreted. A prominent example is the 
Burkhardt-Cottingham sumrule. The fact that the 0th moment does not occur in the Mellin moment decomposition of the polarized 
structure function $g_1(x,Q^2)$ does not mean that the associated integral vanishes as a consequence of the light cone expansion.
In fact, the proof of the Burkhardt-Cottingham sumrule needs quite different techniques
\cite{Burkhardt:1970ti,Jackiw:1972ee}.}

In the calculation of the one-- and two--loop contributions we also used the package {\tt EvaluateMultiSums} 
\cite{EMSSP} and also applied {\tt LiteRed} \cite{LR} for some checks, cf.~\cite{TWOLOOP}. The irreducible three--loop 
diagrams are reduced to 252 master integrals using the 
code {\tt Crusher} \cite{CRUSHER} by applying the integration--by--parts relations \cite{IBP,Chetyrkin:1981qh}. 
Relations between a small 
number of $t$-dependent master integrals are difficult to prove analytically for general values of $D$. However, 
they 
can be proven for the whole finite range of Mellin $N$ and $\ep$ used in the present analysis by the method of 
arbitrary large moments \cite{Blumlein:2017dxp}. For the calculation of the necessary initial values for the difference
equations we use the results given in \cite{Chetyrkin:1981qh,INIT}.

The method of arbitrary large moments implemented within the package 
\texttt{SolveCoupledSystem}~\cite{Blumlein:2019hfc} is also used to generate a
large number of moments for the massless OMEs.
By using the 
method of guessing \cite{GUESS,Blumlein:2009tj} and its implementation in {\tt Sage} 
\cite{SAGE,GSAGE} we determine the difference equations, which correspond to the different color and
multiple zeta value factors \cite{Blumlein:2009cf}. To calculate $\gamma_{\rm NS}^{(2), \pm}$ we generate 
3000 even resp.\ odd moments and for  $\gamma_{\rm NS}^{(2), s}$ 500 moments.
It turns out that the determination of the largest recurrence requires 
1537 moments for $\gamma_{\rm NS}^{(2), +}$,  
1568 moments for $\gamma_{\rm NS}^{(2), -}$, 
1104  moments for $\gamma_{\rm NS}^{(2) \rm +,tr}$ and  
for $\gamma_{\rm NS}^{(2) \rm -,tr}$, and
 348 moments for $\gamma_{\rm NS}^{(2),s}$.
The difference equations are solved by using
methods from difference field theory \cite{DRING} implemented in the package 
{\tt Sigma} \cite{SIG1,SIG2} utilizing functions from {\tt HarmonicSums} 
\cite{HARMSU, Blumlein:2009ta,Vermaseren:1998uu,Blumlein:1998if,Remiddi:1999ew,Ablinger:2011te,Ablinger:2013cf,
Ablinger:2014bra}, to  obtain the three--loop anomalous dimensions. 
The largest difference equation contributing has order 
{\sf o} = {16} and degree  {\sf d} = {304}. Comparing to the reconstruction of the anomalous 
dimensions out of their moments performed in Ref.~\cite{Blumlein:2009tj} the largest 
difference equation had order {\sf o} = 16 and degree {\sf d} = 192, 
requiring 1079 moments.
The overall computation time 
using the automated chain of codes described amounted to about $20$ days
of CPU time on {\tt Intel(R) Xeon(R) CPU E5-2643 v4} processors.
In the present calculation we kept only one power in the gauge parameter $\hat{\xi}$ to check the 
renormalization, which has been sufficient to compute the non--singlet anomalous dimensions. In 
calculating the complete OMEs, no gauge--dependent contribution can be neglected.

The anomalous dimensions, $\gamma_{\rm NS}$, can be expressed by harmonic sums 
\cite{Vermaseren:1998uu,Blumlein:1998if}
\begin{eqnarray}
S_{b,\vec{a}}(N) &=& \sum_{k=1}^N \frac{({\rm sign}(b))^k}{k^{|b|}} S_{\vec{a}}(k),~~~S_\emptyset 
= 
1,~~~b_, a_i \in \mathbb{Z} \backslash \{0\}, N \in \mathbb{N}  \backslash \{0\}. 
\end{eqnarray}
Their Mellin inversion to the splitting functions $P_{qq}(z)$ 
\begin{eqnarray}
\gamma_{qq}(N) = - \int_0^1 dz z^{N-1} P_{qq}(z) 
\end{eqnarray}
can be performed using routines of the packages {\tt HarmonicSums} and is expressed in terms of harmonic polylogarithms 
\cite{Remiddi:1999ew} given by
\begin{eqnarray}
\HA_{b,\vec{a}}(z) &=& \int_0^z dx f_b(x) \HA_{\vec{a}}(x),~~~\HA_\emptyset = 1,~~~b, a_i \in 
\{-1,0,1\},
\end{eqnarray}
with  the  alphabet of letters
\begin{eqnarray}
\mathfrak{A}_{\rm H} = \Biggl\{f_{0}(z) = \frac{1}{z},~~~f_{-1}(z) = \frac{1}{1+z},~~f_{1}(z) = 
\frac{1}{1-z}\Biggr\}.
\end{eqnarray}
In $z$--space one usually distinguishes three contributions to the individual splitting functions, 
because of their different treatment in Mellin convolutions,
\begin{eqnarray}
P(z) = P^\delta(z)  + P^{\rm plu}(z) + P^{\rm reg}(z),
\end{eqnarray}
where $P^\delta(z) = p_0 \delta(1-z)$, $P^{\rm reg}(z)$ is a regular function in $z \in [0,1]$ and 
$P^{\rm plu}(z)$ denotes the remaining genuine $+$-distribution, the Mellin transformation of 
which is given 
by
\begin{eqnarray}
\int_0^1 dz (z^{N-1} - 1) P^{\rm plu}(z).
\end{eqnarray}
We will use this representation in Section~\ref{sec:4}.
\section{The anomalous dimensions and splitting functions} 
\label{sec:4}

\vspace*{1mm} 
\noindent 
In the following we use the minimal representations in terms of the contributing harmonic sums and harmonic 
polylogarithms by applying the algebraic relations between the 
harmonic sums and the harmonic polylogarithms \cite{Blumlein:2003gb}. 26 harmonic sums up to weight {\sf w = 5} contribute.
Both the anomalous dimensions in the vector case, 
$\gamma_{\rm NS}^{(2),\pm}(N)$, and for transversity, $\gamma_{\rm NS}^{(2),\pm, \rm tr}(N)$, are sometimes written
in terms of the difference $\gamma_{\rm NS}^{(2),+}(N) - \gamma_{\rm NS}^{(2),-}(N)$.
This is somewhat problematic, since $\gamma_{\rm NS}^{(2),+}(N)$ 
is defined for positive even integers only, while $\gamma_{\rm NS}^{(2),-}(N)$ refers to positive odd integers.
Later the respective analytic continuations from $N \in \mathbb{N} \to \mathbb{C}$ proceeds 
from the even or the odd integers
\cite{Blumlein:2009ta}.
We will therefore refer to the complete expressions, respectively, as long as they are written in terms of harmonic sums. 
Considering their Mellin inversion to $z$ space
allows then to consider the respective difference term, since the corresponding expression is free of $N$.

We obtain the following expressions for the non--singlet anomalous dimensions in Mellin $N$ space, using
the shorthand notation $S_{\vec{a}}(N) \equiv S_{\vec{a}}$. In the vector case they are given by
\begin{eqnarray}
\label{eq:ga1}
 && \hspace*{-0.7cm} \gamma_{\rm NS}^{(2),+} =
\frac{1}{2} \left[1 + (-1)^N\right] 
\nonumber\\ && \times
\Biggl\{
\textcolor{blue}{C_F^2} 
\Biggl\{
        \textcolor{blue}{C_A} 
\Biggl[
                \frac{72 P_3}{N^2 (1+N)^2} \zeta_3
                +\frac{32 P_{15}}{9 N^2 (1+N)^2} S_{-2,1}
                -\frac{16 P_{17}}{9 N^2 (1+N)^2} S_3
                +\frac{P_{33}}{18 N^4 (1+N)^4}
\nonumber\\ &&    
            +\Biggl(
                        -\frac{16 P_{29}}{9 N^4 (1+N)^4}
                        -\frac{4288}{9} S_2 
                        +\frac{64 \big(
                                -12+31 N+31 N^2\big) }{3 N (1+N)} S_3
                        +320 S_4 
                        -1024 S_{3,1} 
\nonumber\\ &&
                        +\frac{64 \big(
                                -84+31 N+31 N^2\big)  }{3 N (1+N)} S_{-2,1}
                        +3712 S_{-2,2} 
                        +3840 S_{-3,1} 
                        -7168 S_{-2,1,1} 
                \Biggr) S_1 
                +\Biggl(
                        256 S_3 
\nonumber\\ && 
                        +1792 S_{-2,1} 
                \Biggr) S_1 ^2
                +\Biggl(
                        \frac{4 P_{19}}{9 N^2 (1+N)^2}
                        -832 S_3 
                        -5248 S_{-2,1} 
                \Biggr) S_2 
                +\frac{352}{3} S_2 ^2
\nonumber\\ && 
                +\frac{16 \big(
                        -30+151 N+151 N^2\big) }{3 N (1+N)} S_4
                +\Biggl(
                        -\frac{16 P_{22}}{9 N^2 (1+N)^3}
                        +\big(
                                -\frac{64 P_9}{9 N^2 (1+N)^2}
                                -256 S_2 
                        \Biggr) S_1 
\nonumber\\ &&                
         +\frac{32 \big(
                                12+31 N+31 N^2\big) S_2 }{3 N (1+N)}
                        +64 S_3 
                        +5376 S_{2,1} 
                        -384 S_{-2,1} 
                        +576 \zeta_3
                \Biggr) S_{-2} 
\nonumber\\ &&                 
+\Biggl(
                        -\frac{32 \big(
                                8+3 N+3 N^2\big)}{N (1+N)}
                        +512 S_1 
                \Biggr) S_{-2} ^2
                +\Biggl(
                        \frac{32 \big(
                                108+31 N+31 N^2\big)  }{3 N (1+N)} S_1
                        -\frac{16 P_{16}
                        }{9 N^2 (1+N)^2}
\nonumber\\ &&                
         -1152 S_1 ^2
                        +2624 S_2 
                        +960 S_{-2} 
                \Biggr) S_{-3} 
                +\Biggl(
                        \frac{16 \big(
                                138+35 N+35 N^2\big)}{3 N (1+N)}
                        -1472 S_1 
                \Biggr) S_{-4} 
\nonumber\\ &&                
 +2304 S_{-5} 
                +768 S_{2,3} 
                +2688 S_{2,-3} 
                -\frac{64 \big(
                        -24+29 N+29 N^2\big)  }{3 N (1+N)} S_{3,1}
                -768 S_{4,1} 
\nonumber\\ &&                
 +\frac{32 \big(
                        -174+31 N+31 N^2\big) }{3 N (1+N)} S_{-2,2}
                -3648 S_{-2,3} 
                -\frac{1920  }{N (1+N)} S_{-3,1}
                +1728 S_{-4,1} 
\nonumber\\ &&                 
-5376 S_{2,1,-2} 
                +1536 S_{3,1,1} 
                -\frac{128 \big(
                        -84+31 N+31 N^2\big)  }{3 N (1+N)} S_{-2,1,1}
                -1536 S_{-2,1,-2} 
\nonumber\\ &&               
 -5376 S_{-2,2,1} 
                -5376 S_{-3,1,1} 
                +10752 S_{-2,1,1,1} 
\Biggr]
        +\textcolor{blue}{T_F N_F} 
\Biggl[
                -\frac{16   P_5}{9 N^2 (1+N)^2} S_2
                +\frac{4 P_{34}}{9 N^4 (1+N)^4}
\nonumber\\ &&                 
+\Biggl(
                        -\frac{8 P_{13}}{9 N^2 (1+N)^2}
                        +\frac{1280}{9} S_2 
                        -\frac{512}{3} S_3 
                        -\frac{512}{3} S_{-2,1} 
                        +128 \zeta_3
                \Biggr) S_1 
                -\frac{128}{3} S_2 ^2
\nonumber\\ &&           
+\frac{64 \big(
                        12+29 N+29 N^2\big)  }{9 N (1+N)} S_3
                -
                \frac{512}{3} S_4 
                +\Biggl(
                        -\frac{128 \big(
                                -3+10 N+16 N^2\big)}{9 N^2 (1+N)^2}
                        +\frac{2560}{9} S_1 
                        -\frac{256}{3} S_2 
                \Biggr) 
\nonumber\\ &&  \times
S_{-2} 
                +\Biggl(
                        \frac{128 \big(
                                3+10 N+10 N^2\big)}{9 N (1+N)}
                        -\frac{256}{3} S_1 
                \Biggr) S_{-3} 
                -\frac{256}{3} S_{-4} 
                -\frac{256 \big(
                        -3+10 N+10 N^2\big) }{9 N (1+N)} S_{-2,1}
\nonumber\\ && 
                +\frac{256}{3} S_{3,1} 
                -\frac{256}{3} S_{-2,2} 
                +\frac{1024}{3} S_{-2,1,1} 
                -\frac{32 \big(
                        2+3 N+3 N^2\big)}{N (1+N)} \zeta_3
\Biggr]
\Biggr\}
\nonumber\\ && 
+\textcolor{blue}{C_F} 
\Biggl\{
        \textcolor{blue}{T_F^2 N_F^2} 
\Biggl[
                \frac{8 P_{28}}{27 N^3 (1+N)^3}
                -\frac{128}{27} S_1 
                -\frac{640}{27} S_2 
                +\frac{128}{9} S_3 
\Biggr]
        +\textcolor{blue}{C_A^2} 
\Biggl[
                -\frac{24 P_3}{N^2 (1+N)^2} \zeta_3
\nonumber\\ &&                 
-\frac{32  P_{11}}{9 N^2 (1+N)^2} S_{-2,1}
                +\frac{8   P_{18}}{9 N^2 (1+N)^2} S_3
                +\frac{P_{32}}{54 N^3 (1+N)^3}
                +\Biggl(
                        \frac{4 P_{35}}{3 N^4 (1+N)^4}
\nonumber\\ &&                
         -\frac{16 \big(
                                -8+11 N+11 N^2\big) }{N (1+N)} S_3
                        -256 S_4 
                        +512 S_{3,1} 
                        -\frac{64 \big(
                                -24+11 N+11 N^2\big)  }{3 N (1+N)} S_{-2,1}
\nonumber\\ &&                
         -1024 S_{-2,2}
                        -1024 S_{-3,1} 
                        +2048 S_{-2,1,1} 
                \Biggr) S_1 
                +\Biggl(
                        -128 S_3 
                        -512 S_{-2,1} 
                \Biggr) S_1 ^2
                +\Biggl(
                        -
                        \frac{8344}{27}
\nonumber\\ &&               
         +384 S_3 
                        +1536 S_{-2,1} 
                \Biggr) S_2 
                -\frac{16 \big(
                        -24+55 N+55 N^2\big) }{3 N (1+N)} S_4
                +64 S_5 
                +\Biggl(
                        \frac{32   P_{10}}{9 N^2 (1+N)^2} S_1
\nonumber\\ &&                
         +\frac{16 P_{27}}{9 N^3 (1+N)^3}
                        -\frac{352}{3} S_2 
                        -64 S_3 
                        -1536 S_{2,1} 
                        +128 S_{-2,1} 
                        -192 \zeta_3
                \Biggr) S_{-2} 
\nonumber\\ &&                 
+\Biggl(
                        \frac{48 \big(
                                2+N+N^2\big)}{N (1+N)}
                        -192 S_1 
                \Biggr) S_{-2} ^2
                +\Biggl(
                        \frac{16 P_{12}}{9 N^2 (1+N)^2}
                        -\frac{32 \big(
                                24+11 N+11 N^2\big)  }{3 N (1+N)} S_1
\nonumber\\ &&                
         +256 S_1 ^2
                        -768 S_2 
                        -320 S_{-2} 
                \Biggr) S_{-3} 
                +\Biggl(
                        -\frac{16 \big(
                                30+13 N+13 N^2\big)}{3 N (1+N)}
                        +320 S_1 
                \Biggr) S_{-4} 
\nonumber\\ &&                
 -704 S_{-5} 
                -384 S_{2,3} 
                -768 S_{2,-3} 
                +\frac{64 \big(
                        -12+11 N+11 N^2\big)  }{3 N (1+N)} S_{3,1}
                +384 S_{4,1} 
\nonumber\\ &&                
 -\frac{32 \big(
                        -48+11 N+11 N^2\big)  }{3 N (1+N)} S_{-2,2}
                +1088 S_{-2,3} 
                +\frac{512 }{N (1+N)} S_{-3,1}
                -448 S_{-4,1} 
\nonumber\\ &&                
 +1536 S_{2,1,-2} 
                -768 S_{3,1,1} 
                +\frac{128 \big(
                        -24+11 N+11 N^2\big)  }{3 N (1+N)} S_{-2,1,1}
                +512 S_{-2,1,-2} 
                +1536 [S_{-2,2,1} 
\nonumber\\ &&                
 +S_{-3,1,1}] 
                -3072 S_{-2,1,1,1} 
        \Biggr]
        +\textcolor{blue}{C_A T_F N_F} \Biggl[
                -
                \frac{8 P_{31}}{27 N^3 (1+N)^3}
                +\Biggl(
                        -\frac{16 P_{30}}{27 N^3 (1+N)^3}
                        +64 S_3 
\nonumber\\ &&                
         +\frac{256}{3} S_{-2,1} 
                        -128 \zeta_3
                \Biggr) S_1 
                +\frac{5344}{27} S_2 
                -\frac{32 \big(
                        3+14 N+14 N^2\big)}{3 N (1+N)} S_3
                +\frac{320}{3} S_4 
                +\Biggl(
                        -\frac{1280}{9} S_1 
\nonumber\\ && 
                       + \frac{64 \big(
                                -3+10 N+16 N^2\big)}{9 N^2 (1+N)^2}
                        +\frac{128}{3} S_2 
                \Biggr) S_{-2} 
                +\Biggl(
                        -\frac{64 \big(
                                3+10 N+10 N^2\big)}{9 N (1+N)}
                        +\frac{128}{3} S_1 
                \Biggr) S_{-3} 
\nonumber\\ &&                
                +\frac{128}{3} S_{-4} 
 -\frac{256}{3} S_{3,1} 
                +\frac{128 \big(
                        -3+10 N+10 N^2\big) }{9 N (1+N)} S_{-2,1}
                +\frac{128}{3} S_{-2,2} 
                -\frac{512}{3} S_{-2,1,1} 
\nonumber\\ &&                
 +\frac{32 \big(
                        2+3 N+3 N^2\big)}{N (1+N)} \zeta_3
\Biggl]
\Biggr\}
+\textcolor{blue}{C_F^3} \Biggl\{
        -\frac{48 P_3}{N^2 (1+N)^2} \zeta_3
        +\frac{8 P_4}{N^2 (1+N)^2} S_3 
        +\frac{P_{36}}{N^5 (1+N)^5}
\nonumber\\ &&         
+\Biggl(
                \frac{8 P_{26}}{N^4 (1+N)^4}
                -\frac{128 (1+2 N) }{N^2 (1+N)^2} S_2
                +128 S_2 ^2
                -384 S_3 
                +128 S_4 
                +512 S_{3,1} 
                -3328 S_{-2,2} 
\nonumber\\ && 
                -\frac{384 \big(
                        -4+N+N^2\big) }{N (1+N)} S_{-2,1}
                -3584 S_{-3,1} 
                +6144 S_{-2,1,1} 
        \Biggr) S_1 
        +\Biggl(
                -
                \frac{64 \big(
                        1+3 N+3 N^2\big)}{N^3 (1+N)^3}
\nonumber\\ &&                
 -1536 S_{-2,1} 
        \Biggr) S_1 ^2
        +\Biggl(
                \frac{4 P_{25}}{N^3 (1+N)^3}
                +512 S_3 
                +4352 S_{-2,1} 
        \Biggr) S_2 
        -\frac{32 \big(
                2+3 N+3 N^2\big)}{N (1+N)} S_2^2
\nonumber\\ &&         
-\frac{32 \big(
                2+15 N+15 N^2\big) }{N (1+N)} S_4
        +\big(
                \frac{32 P_{24}}{N^3 (1+N)^3}
                +\Biggl(
                        -\frac{128 \big(
                                5+7 N+3 N^2\big)}{N^2 (1+N)^2}
                        +512 S_2 
                \Biggr) S_1 
\nonumber\\ &&                
 -\frac{64 \big(
                        4+3 N+3 N^2\big)  }{N (1+N)} S_2
                +128 S_3 
                -4608 S_{2,1} 
                +256 S_{-2,1} 
                -384 \zeta_3
        \Biggr) S_{-2} 
        +\Biggl(
                \frac{128}{N (1+N)}
\nonumber\\ &&                
 -256 S_1 
        \Biggr) S_{-2} ^2
        +\Biggl(
                \frac{32 \big(
                        8+5 N+9 N^2\big)}{N^2 (1+N)^2}
                -\frac{64 \big(
                        20+3 N+3 N^2\big) }{N (1+N)} S_1
                +1280 S_1 ^2
                -2176 S_2 
\nonumber\\ &&                
 -640 S_{-2} 
        \Biggr) S_{-3} 
        +\Biggl(
                -\frac{32 \big(
                        26+3 N+3 N^2\big)}{N (1+N)}
                +1664 S_1 
        \Biggr) S_{-4} 
        -1792 S_{-5} 
        -384 S_{2,3} 
\nonumber\\ && 
        -2304 S_{2,-3} 
        +\frac{128 \big(
                -2+3 N+3 N^2\big)  }{N (1+N)} S_{3,1}
        +384 S_{4,1} 
        -\frac{64 \big(
                4-N+3 N^2\big) }{N^2 (1+N)^2} S_{-2,1}
\nonumber\\ &&         
-
        \frac{64 \big(
                -26+3 N+3 N^2\big)  }{N (1+N)} S_{-2,2}
        +2944 S_{-2,3} 
        +\frac{1792 }{N (1+N)} S_{-3,1}
        -1664 S_{-4,1} 
\nonumber\\ &&         
+4608 S_{2,1,-2} 
        -768 S_{3,1,1} 
        +\frac{768 \big(
                -4+N+N^2\big) }{N (1+N)} S_{-2,1,1}
        +1024 S_{-2,1,-2} 
\nonumber\\ && 
        +4608 [S_{-2,2,1} + S_{-3,1,1}] 
        -9216 S_{-2,1,1,1} 
\Biggr\}
\Biggr\}, 

\\
\label{eq:ga2}
\hspace*{-1cm}
&& \gamma_{\rm NS}^{(2),-} =
\frac{1}{2} \left[1 - (-1)^N\right]
\nonumber\\ && \times
\Biggl\{\textcolor{blue}{C_F^2} \Biggl\{
        \textcolor{blue}{C_A} \Bigl[
 \frac{16 \big(-126 + 6 N + 427 N^2 + 770 N^3 + 385 N^4\big)}{9 N^2 (1 + N)^2} [1 - S_3]
+\frac{72 \zeta_3 P_{37}}{N^2 (1+N)^2}
\nonumber\\ &&
                +\frac{32 P_{43}}{9 N^2 (1+N)^2} S_{-2,1}
                -\frac{16  P_{17}}{9 N^2 (1+N)^2}
                +\frac{P_{57}}{18 N^5 (1+N)^5}
                +\Biggl(
                        -\frac{16 P_{50}}{9 N^4 (1+N)^4}
-\frac{4288}{9} S_2
\nonumber\\ &&
                        +\frac{64 \big(
                                -12+31 N+31 N^2\big)}{3 N (1+N)} S_3
                        +320 S_4
                        -1024 S_{3,1}
                        +\frac{64 \big(
                                -84+31 N+31 N^2\big)}{3 N (1+N)} S_{-2,1}
\nonumber\\ &&                        
+3712 S_{-2,2}
                        +3840 S_{-3,1}
                        -7168 S_{-2,1,1}
                \Biggr) S_1
                +\Biggl(
                        256 S_3
                        +1792 S_{-2,1}
                \Biggr) S_1^2
                +\Biggl(
                        \frac{4 P_{53}}{9 N^3 (1+N)^3}
\nonumber\\ &&                        
-832 S_3
                        -5248 S_{-2,1}
                \Biggr) S_2
                +\frac{352}{3} S_2^2
                +\frac{16 \big(
                        -30+151 N+151 N^2\big)}{3 N (1+N)} S_4
                +\Biggl(
                        -\frac{16 P_{45}}{9 N^3 (1+N)^2}
\nonumber\\ &&                        
+\Biggl(
                                -\frac{64 P_{40}}{9 N^2 (1+N)^2}
                                -256 S_2
                        \Biggr) S_1
                        +\frac{32 \big(
                                12+31 N+31 N^2\big)}{3 N (1+N)} S_2
                        +64 S_3
                        +5376 S_{2,1}
                        -384 S_{-2,1}
\nonumber\\ &&                        
+576 \zeta_3 
                \Biggr) S_{-2}
                +\Biggl(
                        -\frac{32 \big(
                                8+3 N+3 N^2\big)}{N (1+N)}
                        +512 S_1
                \Biggr) S_{-2}^2
                +\Biggl(
                        \frac{32 \big(
                                108+31 N+31 N^2\big)}{3 N (1+N)} S_1
\nonumber\\ &&                        
-\frac{16 P_{44}}{9 N^2 (1+N)^2}
                        -1152 S_1^2
                        +2624 S_2
                        +960 S_{-2}
                \Biggr) S_{-3}
                +\Biggl(
                        \frac{16 \big(
                                138+35 N+35 N^2\big)}{3 N (1+N)}
                        -1472 S_1
                \Biggr) 
\nonumber\\ && \times
S_{-4}
                +2304 S_{-5}
                +768 S_{2,3}
                +2688 S_{2,-3}
                -\frac{64 \big(
                        -24+29 N+29 N^2\big)}{3 N (1+N)} S_{3,1}
                -768 S_{4,1}
\nonumber\\ &&                
+\frac{32 \big(
                        -174+31 N+31 N^2\big)}{3 N (1+N)} S_{-2,2}
                -3648 S_{-2,3}
                -\frac{1920 S_{-3,1}}{N (1+N)}
                +1728 S_{-4,1}
                -5376 S_{2,1,-2}
\nonumber\\ &&                
+1536 S_{3,1,1}
                -\frac{128 \big(
                        -84+31 N+31 N^2\big)}{3 N (1+N)} S_{-2,1,1}
                -1536 S_{-2,1,-2}
                -5376 S_{-2,2,1}
                -5376 S_{-3,1,1}
\nonumber\\ &&  
               +10752 S_{-2,1,1,1}
        \Bigg]
        +\textcolor{blue}{T_F N_F} \Biggl[
                -\frac{16P_5}{9 N^2 (1+N)^2} S_2
                +\frac{4 P_{54}}{9 N^4 (1+N)^4}
                +\Biggl(
                        -\frac{8 P_{51}}{9 N^3 (1+N)^3}
                        +\frac{1280}{9} S_2
\nonumber\\ &&                         
-\frac{512}{3} S_3
                        -\frac{512}{3} S_{-2,1}
                        +128 \zeta_3
                \Biggr) S_1
                -\frac{128}{3} S_2^2
                +\frac{64 \big(
                        12+29 N+29 N^2\big)}{9 N (1+N)} S_3
                -
                \frac{512}{3} S_4
\nonumber\\ &&                
+\Biggl(
                        -\frac{128 \big(
                                -3+10 N+16 N^2\big)}{9 N^2 (1+N)^2}
                        +\frac{2560}{9} S_1
                        -\frac{256}{3} S_2
                \Biggr) S_{-2}
                +\Biggl(
                        \frac{128 \big(
                                3+10 N+10 N^2\big)}{9 N (1+N)}
\nonumber\\ &&                         
-\frac{256}{3} S_1
                \Biggr) S_{-3}
                -\frac{256}{3} S_{-4}
                +\frac{256}{3} S_{3,1}
                -\frac{256 \big(
                        -3+10 N+10 N^2\big)}{9 N (1+N)} S_{-2,1}
                -\frac{256}{3} S_{-2,2}
\nonumber\\ &&                
 +\frac{1024}{3} S_{-2,1,1}
                -\frac{32 \big(
                        2+3 N+3 N^2\big)}{N (1+N)} \zeta_3
        \Biggr]
\Biggr\}
+\textcolor{blue}{C_F} \Biggl\{
        \textcolor{blue}{T_F^2 N_F^2} \Biggl[
                \frac{8 P_{28}}{27 N^3 (1+N)^3}
                -\frac{128}{27} S_1
\nonumber\\ &&                
 -\frac{640}{27} S_2
                +\frac{128}{9} S_3
        \Biggr]
        +\textcolor{blue}{C_A^2} \Biggl[
                -\frac{24 P_{37}}{N^2 (1+N)^2} \zeta_3
                -\frac{32 P_{41}}{9 N^2 (1+N)^2} S_{-2,1}
                +\frac{8 P_{18}}{9 N^2 (1+N)^2} S_3
\nonumber\\ &&                 
+\frac{P_{59}}{54 N^5 (1+N)^5}
                +\Biggl(
                        \frac{4 P_{55}}{3 N^4 (1+N)^4}
                        -\frac{16 \big(
                                -8+11 N+11 N^2\big)}{N (1+N)} S_3
                        -256 S_4
                        +512 S_{3,1}
\nonumber\\ &&                
         -\frac{64 \big(
                                -24+11 N+11 N^2\big)}{3 N (1+N)} S_{-2,1}
                        -1024 S_{-2,2}
                        -1024 S_{-3,1}
                        +2048 S_{-2,1,1}
                \Biggr) S_1
\nonumber\\ &&                
 +\Biggl(
                        -128 S_3
                        -512 S_{-2,1}
                \Biggr) S_1^2
                +\Biggl(
                        -
                        \frac{8344}{27}
                        +384 S_3
                        +1536 S_{-2,1}
                \Biggr) S_2
                +64 S_5
\nonumber\\ &&                
 -\frac{16 \big(
                        -24+55 N+55 N^2\big)}{3 N (1+N)} S_4
                +\Biggl(
                        \frac{32  P_{39}}{9 N^2 (1+N)^2} S_1
                        +\frac{16 P_{49}}{9 N^3 (1+N)^3}
                        -\frac{352}{3} S_2
                        -64 S_3
\nonumber\\ && 
         -1536 S_{2,1}
                        +128 S_{-2,1}
                        -192 \zeta_3
                \Biggr) S_{-2}
                +\Biggl(
                        \frac{48 \big(
                                2+N+N^2\big)}{N (1+N)}
                        -192 S_1
                \Biggr) S_{-2}^2
                +\Biggl(
                        \frac{16 P_{42}}{9 N^2 (1+N)^2}
\nonumber\\ &&                         
-\frac{32 \big(
                                24+11 N+11 N^2\big)}{3 N (1+N)} S_1
                        +256 S_1^2
                        -768 S_2
                        -320 S_{-2}
                \Biggr) S_{-3}
                +\Biggl(
                        -\frac{16 \big(
                                30+13 N+13 N^2\big)}{3 N (1+N)}
\nonumber\\ &&                
         +320 S_1
                \Biggr) S_{-4}
                -704 S_{-5}
                -384 S_{2,3}
                -768 S_{2,-3}
                +\frac{64 \big(
                        -12+11 N+11 N^2\big)}{3 N (1+N)} S_{3,1}
\nonumber\\ &&                
 +384 S_{4,1}
                -\frac{32 \big(
                        -48+11 N+11 N^2\big) S_{-2,2}}{3 N (1+N)}
                +1088 S_{-2,3}
                +\frac{512 S_{-3,1}}{N (1+N)}
                -448 S_{-4,1}
\nonumber\\ &&                
 +1536 S_{2,1,-2}
                -768 S_{3,1,1}
                +\frac{128 \big(
                        -24+11 N+11 N^2\big)}{3 N (1+N)} S_{-2,1,1}
                +512 S_{-2,1,-2}
                +1536 [S_{-2,2,1}
\nonumber\\ && 
                +  S_{-3,1,1}]
                -3072 S_{-2,1,1,1}
        \Biggr]
        +\textcolor{blue}{C_A T_F N_F}\Biggl[
                -
                \frac{8 P_{56}}{27 N^4 (1+N)^4}
                +\Biggl(
                        -\frac{16 P_{52}}{27 N^3 (1+N)^3}
                        +64 S_3
\nonumber\\ && 
                        +\frac{256}{3} S_{-2,1}
                        -128 \zeta_3
                \Biggr) S_1
                +\frac{5344}{27} S_2
                -\frac{32 \big(
                        3+14 N+14 N^2\big)}{3 N (1+N)} S_3
                +\frac{320}{3} S_4
\nonumber\\ &&                
 +\Biggl(
                        \frac{64 \big(
                                -3+10 N+16 N^2\big)}{9 N^2 (1+N)^2}
                        -\frac{1280}{9} S_1
                        +\frac{128}{3} S_2
                \Biggr) S_{-2}
                +\Biggl(
                        -\frac{64 \big(
                                3+10 N+10 N^2\big)}{9 N (1+N)}
\nonumber\\ &&                
         +\frac{128}{3} S_1
                \Biggr) S_{-3}
                +\frac{128}{3} S_{-4}
                -\frac{256}{3} S_{3,1}
                +\frac{128 \big(
                        -3+10 N+10 N^2\big)}{9 N (1+N)} S_{-2,1}
                +\frac{128}{3} S_{-2,2}
\nonumber\\ &&                
 -\frac{512}{3} S_{-2,1,1}
                +\frac{32 \big(
                        2+3 N+3 N^2\big)}{N (1+N)} \zeta_3
        \Biggr]
\Biggr\}
+\textcolor{blue}{C_F^3} \Biggl\{
        -\frac{48 \zeta_3 P_{37}}{N^2 (1+N)^2}
        +\frac{8 P_{38}}{N^2 (1+N)^2} S_3
\nonumber\\ &&         
+\frac{P_{58}}{N^5 (1+N)^5}
        +\Biggl(
                \frac{8 P_{48}}{N^4 (1+N)^4}
                -\frac{128 (1+2 N)}{N^2 (1+N)^2} S_2
                +128 S_2^2
                -384 S_3
                +128 S_4
                +512 S_{3,1}
\nonumber\\ &&                
 -\frac{384 \big(
                        -4+N+N^2\big)}{N (1+N)} S_{-2,1}
                -3328 S_{-2,2}
                -3584 S_{-3,1}
                +6144 S_{-2,1,1}
        \Biggr) S_1
\nonumber\\ &&         
+\Biggl(
                -
                \frac{64 \big(
                        1+3 N+3 N^2\big)}{N^3 (1+N)^3}
                -1536 S_{-2,1}
        \Biggr) S_1^2
        +\Biggl( 
                \frac{4 P_{47}}{N^3 (1+N)^3}
                +512 S_3
                +4352 S_{-2,1}
        \Biggr) S_2
\nonumber\\ &&         
-\frac{32 \big(
                2+3 N+3 N^2\big)}{N (1+N)} S_2^2
        -\frac{32 \big(
                2+15 N+15 N^2\big)}{N (1+N)} S_4
        +\Biggl(
                \frac{32 P_{46}}{N^3 (1+N)^3}
                        +512 S_2
\nonumber\\ && 
                +\Biggl(
                        \frac{128 \big(
                                1-N+3 N^2\big)}{N^2 (1+N)^2}
+ 512 S_2                 
\Biggl) S_1
                -\frac{64 \big(
                        4+3 N+3 N^2\big)}{N (1+N)} S_2 - 512  S_2 
                +128 S_3
                -4608 S_{2,1}
\nonumber\\ &&                
                +256 S_{-2,1}
 -384 \zeta_3
        \Biggr) S_{-2}
        +\Biggl(
                \frac{128}{N (1+N)}
                -256 S_1
        \Biggr) S_{-2}^2
        +\Biggl(
                \frac{32 \big(
                        20+17 N+21 N^2\big)}{N^2 (1+N)^2}
 +1280 S_1^2
\nonumber\\ &&                
                -2176 S_2
                -\frac{64 \big(
                        20+3 N+3 N^2\big) }{N (1+N)} S_1
                -640 S_{-2}
        \Biggr) S_{-3}
        +\Biggl(
                -\frac{32 \big(
                        26+3 N+3 N^2\big)}{N (1+N)}
                +1664 S_1
        \Biggr) S_{-4}
\nonumber\\ &&         
        -1792 S_{-5}
-384 S_{2,3}
        -2304 S_{2,-3}
        +\frac{128 \big(
                -2+3 N+3 N^2\big)}{N (1+N)} S_{3,1}
        +384 S_{4,1}
        -1664 S_{-4,1}
\nonumber\\ && 
        +4608 S_{2,1,-2}
        -\frac{64 \big(
                16+11 N+15 N^2\big) 
        }{N^2 (1+N)^2} S_{-2,1}
        -\frac{64 \big(
                -26+3 N+3 N^2\big)}{N (1+N)} S_{-2,2}
        +2944 S_{-2,3}
\nonumber\\ &&        
 +\frac{1792 }{N (1+N)} S_{-3,1}
        -768 S_{3,1,1}
        +\frac{768 \big(
                -4+N+N^2\big)}{N (1+N)} S_{-2,1,1}
        +1024 S_{-2,1,-2}
        +4608 [S_{-2,2,1} 
\nonumber\\ &&        
+ S_{-3,1,1}]
        -9216 S_{-2,1,1,1}
\Biggr\}
\Biggr\}.

\end{eqnarray}
In the case of transversity we obtain 
\begin{eqnarray}
\label{eq:ga3}
&& \hspace*{-7mm} \gamma_{\rm NS}^{(2), \rm tr,+} = \frac{1}{2}\left[1 + (-1)^N\right] \nonumber\\ && \times 
\Biggl\{
\textcolor{blue}{C_F} \Biggl\{
        \textcolor{blue}{T_F^2 N_F^2} \Biggl[
                \frac{8 \big(
                        -8+17 N+17 N^2\big)}{9 N (1+N)}
                -\frac{128}{27} S_1 
                -\frac{640}{27} S_2 
                +\frac{128}{9} S_3 
        \Biggr]
\nonumber\\ && 
        +\textcolor{blue}{C_A T_F N_F} \Biggl[
                -
                \frac{16 \big(
                        -22+45 N+45 N^2\big)}{9 N (1+N)}
                +\Biggl(
                        -\frac{16 \big(
                                9+209 N+209 N^2\big)}{27 N (1+N)}
                        +64 S_3 
\nonumber\\ &&                  
       +\frac{256}{3} S_{-2,1} 
                        -128 \zeta_3
                \Biggr) S_1 
                +\frac{5344}{27} S_2 
                -\frac{448}{3} S_3 
                +\frac{320}{3} S_4 
                +\Biggl(
                        -\frac{1280}{9} S_1 
                        +\frac{128}{3} S_2 
                \Biggr) S_{-2} 
\nonumber\\ &&                
 +\Biggl(
                        -\frac{640}{9}
                        +\frac{128}{3} S_1 
                \Biggr) S_{-3} 
                +\frac{128}{3} S_{-4} 
                -\frac{256}{3} S_{3,1} 
                +\frac{1280}{9} S_{-2,1} 
                +\frac{128}{3} S_{-2,2} 
                -\frac{512}{3} S_{-2,1,1} 
\nonumber\\ &&                
 +96 \zeta_3
        \Biggr]
        +\textcolor{blue}{C_A^2} \Biggl[
                \frac{-968+1657 N+1657 N^2}{18 N (1+N)}
                +\Biggl(
                        \frac{4 P_{14}}{3 (-1+N) N (1+N) (2+N)}
                        -176 S_3 
\nonumber\\ &&                
         -256 S_4 
                        +512 S_{3,1} 
                        -\frac{704}{3} S_{-2,1} 
                        -1024 S_{-2,2} 
                        -1024 S_{-3,1} 
                        +2048 S_{-2,1,1} 
                \Biggr) S_1 
                +\Biggl(
                        -128 S_3 
\nonumber\\ &&                
         -512 S_{-2,1} 
                \Biggr) S_1 ^2
                +\Biggl(
                        -\frac{8344}{27}
                        +384 S_3 
                        +1536 S_{-2,1} 
                \Biggr) S_2 
                +\frac{3112}{9} S_3 
                -\frac{880}{3} S_4 
                +64 S_5 
\nonumber\\ &&                
 +\Biggl(
                        \frac{16 P_2}{(-1+N) N (1+N) (2+N)}
                        +
                        \frac{32 \big(
                                -241+134 N+134 N^2\big) S_1 }{9 (-1+N) (2+N)}
                        -\frac{352}{3} S_2 
                        -64 S_3 
\nonumber\\ &&                
         -1536 S_{2,1} 
                        +128 S_{-2,1} 
                        -192 \zeta_3
                \Biggr) S_{-2} 
                +\Biggl(
                        48
                        -192 S_1 
                \Biggr) S_{-2} ^2
                +\Biggl(
                        256 S_1 ^2
                        -768 S_2 
                        -320 S_{-2} 
\nonumber\\ &&                 
+        \frac{32 \big(
                                -107+67 N+67 N^2\big)}{9 (-1+N) (2+N)}
                        -\frac{352}{3} S_1 
                \Biggr) S_{-3} 
                +\Biggl(
                        -\frac{208}{3}
                        +320 S_1 
                \Biggr) S_{-4} 
                -704 S_{-5} 
                -384 S_{2,3} 
\nonumber\\ &&                
 -768 S_{2,-3} 
                +\frac{704}{3} S_{3,1} 
                +384 S_{4,1} 
                -\frac{64 \big(
                        -107+67 N+67 N^2\big) S_{-2,1} }{9 (-1+N) (2+N)}
                -\frac{352}{3} S_{-2,2} 
\nonumber\\ &&                
 +1088 S_{-2,3} 
                -448 S_{-4,1} 
                +1536 [S_{2,1,-2} 
                + S_{-2,2,1} 
                + S_{-3,1,1}] 
                -768 S_{3,1,1} 
\nonumber\\ &&                 
+\frac{1408}{3} S_{-2,1,1} 
                +512 S_{-2,1,-2} 
                -3072 S_{-2,1,1,1} 
                -\frac{24 \big(
                        -6+5 N+5 N^2\big) \zeta_3}{(-1+N) (2+N)}
        \Biggr]
\Biggr\}
\nonumber\\ && 
+\textcolor{blue}{C_F^2} \Biggl\{
        \textcolor{blue}{T_F N_F} \Biggl[
                92
                +\Biggl(
                        -\frac{8 \big(
                                -8+55 N+55 N^2\big)}{3 N (1+N)}
                        +\frac{1280}{9} S_2 
                        -\frac{512}{3} S_3 
                        -
                        \frac{512}{3} S_{-2,1} 
                        +128 \zeta_3
                \Biggr) 
\nonumber\\ &&  \times
S_1 
                -\frac{80}{3} S_2 
                -\frac{128}{3} S_2 ^2
                +\frac{1856}{9} S_3 
                -\frac{512}{3} S_4 
                +\Biggl(
                        \frac{2560}{9} S_1 
                        -\frac{256}{3} S_2 
                \Biggr) S_{-2} 
                +\Biggl(
                        \frac{1280}{9}
\nonumber\\ &&                
         -\frac{256}{3} S_1 
                \Biggr) S_{-3} 
                -\frac{256}{3} S_{-4} 
                +\frac{256}{3} S_{3,1} 
                -\frac{2560}{9} S_{-2,1} 
                -\frac{256}{3} S_{-2,2} 
                +\frac{1024}{3} S_{-2,1,1} 
                -96 \zeta_3
        \Biggr]
\nonumber\\ && 
        +\textcolor{blue}{C_A} \Biggl[
                -\frac{151}{2}
                +\Biggl(
                        -\frac{8 \big(
                                -206+211 N+211 N^2\big)}{3 (-1+N) N (1+N) (2+N)}
                        -\frac{4288}{9} S_2 
                        +\frac{1984}{3} S_3 
                        +320 S_4 
                        -1024 S_{3,1} 
\nonumber\\ &&                
         +\frac{1984}{3} S_{-2,1} 
                        +3712 S_{-2,2} 
                        +3840 S_{-3,1} 
                        -7168 S_{-2,1,1} 
                \Biggr) S_1 
                +\Biggl(
                        256 S_3 
                        +1792 S_{-2,1} 
                \Biggr) S_1 ^2
\nonumber\\ &&                
 +\Biggl(
                        \frac{604}{3}
                        -832 S_3 
                        -5248 S_{-2,1} 
                \Biggr) S_2 
                +\frac{352}{3} S_2 ^2
                -\frac{6160}{9} S_3 
                +\frac{2416}{3} S_4 
\nonumber\\ &&                
 +\Biggl(
                        -\frac{48 P_2}{(-1+N) N (1+N) (2+N)}
                        +\Biggl(
                                -\frac{64 P_7}{9 (-1+N) N (1+N) (2+N)}
                                -256 S_2 
                        \Biggr) S_1 
\nonumber\\ &&                
         +
                        \frac{992}{3} S_2 
                        +64 S_3 
                        +5376 S_{2,1} 
                        -384 S_{-2,1} 
                        +576 \zeta_3
                \Biggr) S_{-2} 
                +\Biggl(
                        -96
                        +512 S_1 
                \Biggr) S_{-2} ^2
\nonumber\\ &&                
 +\Biggl(
                        -\frac{32 \big(
                                -187+134 N+134 N^2\big)}{9 (-1+N) (2+N)}
                        +\frac{992}{3} S_1 
                        -1152 S_1 ^2
                        +2624 S_2 
                        +960 S_{-2} 
                \Biggr) S_{-3} 
\nonumber\\ &&                
 +\Biggl(
                        \frac{560}{3}
                        -1472 S_1 
                \Biggr) S_{-4} 
                +2304 S_{-5} 
                +768 S_{2,3} 
                +2688 S_{2,-3} 
                -\frac{1856}{3} S_{3,1} 
                -768 S_{4,1} 
\nonumber\\ && 
                +\frac{64 \big(
                        -187+134 N+134 N^2\big) S_{-2,1} }{9 (-1+N) (2+N)}
                +\frac{992}{3} S_{-2,2} 
                -3648 S_{-2,3} 
                +1728 S_{-4,1} 
\nonumber\\ &&                 
-5376 [S_{2,1,-2} 
                + S_{-2,2,1} 
                + S_{-3,1,1}] 
                +1536 [S_{3,1,1} 
                - S_{-2,1,-2} ]
                -\frac{3968}{3} S_{-2,1,1} 
\nonumber\\ &&                 
+10752 S_{-2,1,1,1} 
                +\frac{72 \big(
                        -6+5 N+5 N^2\big) \zeta_3}{(-1+N) (2+N)}
        \Biggr]
\Biggr\}
\nonumber\\ && 
+\textcolor{blue}{C_F^3} \Biggl\{
        -29
        +\Biggl(
                \frac{384 \big(
                        -1+N+N^2\big)}{(-1+N) N (1+N) (2+N)}
                +128 S_2 ^2
                -384 S_3 
                +128 S_4 
                +512 S_{3,1} 
\nonumber\\ &&                
 -384 S_{-2,1} 
                -3328 S_{-2,2}
                                -3584 S_{-3,1} 
                +6144 S_{-2,1,1} 
        \Biggr) S_1 
        -256 S_{-2} ^2 S_1 
        +\Biggl(
                12
                +512 S_3 
\nonumber\\ &&                
 +4352 S_{-2,1} 
        \Biggr) S_2 
        -96 S_2 ^2
        +104 S_3 
        -480 S_4 
        +\Biggl(
                \frac{32 P_2}{(-1+N) N (1+N) (2+N)}
\nonumber\\ &&                
 +\Biggl(
                        \frac{384}{(-1+N) N (1+N) (2+N)}
                        +512 S_2 
                \Biggr) S_1 
                -192 S_2 
                +128 S_3 
                -4608 S_{2,1} 
                +256 S_{-2,1} 
\nonumber\\ &&                
 -384 \zeta_3
        \Biggr) S_{-2} 
        +\Biggl(
                \frac{192}{(-1+N) (2+N)}
                -192 S_1 
                +1280 S_1 ^2
                -2176 S_2 
                -640 S_{-2} 
        \Biggr) S_{-3} 
\nonumber\\ &&         
+\Biggl(
                -96
                +1664 S_1 
        \Biggr) S_{-4} 
        -1792 S_{-5} 
        +384 [-S_{2,3} 
        + S_{3,1} 
        + S_{4,1}] 
        -2304 S_{2,-3} 
\nonumber\\ &&        
-\frac{384 S_{-2,1} }{(-1+N) (2+N)}
        -1536 S_1 ^2 S_{-2,1} 
        -192 S_{-2,2} 
        +2944 S_{-2,3} 
        -1664 S_{-4,1} 
        +4608 S_{2,1,-2} 
\nonumber\\ &&         
-768 S_{3,1,1} 
        +768 S_{-2,1,1} 
        +1024 S_{-2,1,-2} 
        +4608 [S_{-2,2,1} 
        + S_{-3,1,1}] 
        -9216 S_{-2,1,1,1} 
\nonumber\\ &&         
-\frac{48 \big(
                -6+5 N+5 N^2\big) \zeta_3}{(-1+N) (2+N)}
\Biggr\}
\Biggr\},

\\
\label{eq:ga4}
&& \gamma_{\rm NS}^{(2), \rm tr,-} = \frac{1}{2}\left[1 - (-1)^N\right] 
\Biggl\{ 
\textcolor{blue}{C_F} \Biggl\{
        \textcolor{blue}{T_F^2 N_F^2} \Biggl[
                \frac{8 \big(
                        -8+17 N+17 N^2\big)}{9 N (1+N)}
                -\frac{128}{27} S_1 
                -\frac{640}{27} S_2 
                +\frac{128}{9} S_3 
        \Biggr]
\nonumber\\ && 
        +\textcolor{blue}{C_A T_F N_F} \Biggl[
                -\frac{16 \big(
                        -8+49 N+90 N^2+45 N^3\big)}{9 N (1+N)^2}
                +\Biggl(
                        -
                        \frac{16 \big(
                                -27+209 N+209 N^2\big)}{27 N (1+N)}
\nonumber\\ &&                 
        +64 S_3 
                        +\frac{256}{3} S_{-2,1} 
                        -128 \zeta_3
                \Biggr) S_1 
                +\frac{5344}{27} S_2 
                -\frac{448}{3} S_3 
                +\frac{320}{3} S_4 
                +\Biggl(
                        -\frac{1280}{9} S_1 
                        +\frac{128}{3} S_2 
                \Biggr) S_{-2} 
\nonumber\\ &&                
 +\Biggl(
                        -\frac{640}{9}
                        +\frac{128}{3} S_1 
                \Biggr) S_{-3} 
                +\frac{128}{3} S_{-4} 
                -\frac{256}{3} S_{3,1} 
                +\frac{1280}{9} S_{-2,1} 
                +\frac{128}{3} S_{-2,2} 
                -\frac{512}{3} S_{-2,1,1} 
\nonumber\\ &&                 
+96 \zeta_3
        \Biggr]
        +\textcolor{blue}{C_A^2} \Biggl[
                \frac{P_{23}}{18 (-1+N) N (1+N)^2 (2+N)}
                +\Biggl(
                        \frac{4 \big(
                                12+245 N+245 N^2\big)}{3 N (1+N)}
                        -176 S_3 
\nonumber\\ &&                
         -256 S_4 
                        +512 S_{3,1} 
                        -\frac{704}{3} S_{-2,1} 
                        -1024 S_{-2,2} 
                        -1024 S_{-3,1} 
                        +2048 S_{-2,1,1} 
                \Biggr) S_1 
                +\Biggl(
                        -128 S_3 
\nonumber\\ &&                
         -512 S_{-2,1} 
                \Biggr) S_1 ^2
                +\Biggl(
                        -\frac{8344}{27}
                        +384 S_3 
                        +1536 S_{-2,1} 
                \Biggr) S_2 
                +\frac{3112}{9} S_3 
                -\frac{880}{3} S_4 
                +64 S_5 
\nonumber\\ &&                
 +\Biggl(
                        \frac{16 \big(
                                -5+3 N+3 N^2\big)}{(-1+N) (2+N)}
                        +\frac{32 \big(
                                81+134 N+134 N^2\big)}{9 N (1+N)} S_1
                        -
                        \frac{352}{3} S_2 
                        -64 S_3 
                        -1536 S_{2,1} 
\nonumber\\ &&                
         +128 S_{-2,1} 
                        -192 \zeta_3
                \Biggr) S_{-2} 
                +\Biggl(
                        48
                        -192 S_1 
                \Biggr) S_{-2} ^2
                +\Biggl(
                        \frac{32 \big(
                                81+67 N+67 N^2\big)}{9 N (1+N)}
                        -\frac{352}{3} S_1 
\nonumber\\ &&                
         +256 S_1 ^2
                        -768 S_2 
                        -320 S_{-2} 
                \Biggr) S_{-3} 
                +\Biggl(
                        -\frac{208}{3}
                        +320 S_1 
                \Biggr) S_{-4} 
                -704 S_{-5} 
                -384 S_{2,3} 
                -768 S_{2,-3} 
\nonumber\\ &&                
 +\frac{704}{3} S_{3,1} 
                +384 S_{4,1} 
                -\frac{64 \big(
                        81+67 N+67 N^2\big) S_{-2,1} }{9 N (1+N)}
                -\frac{352}{3} S_{-2,2} 
                +1088 S_{-2,3} 
                -448 S_{-4,1} 
\nonumber\\ &&                
 +1536 S_{2,1,-2} 
                -768 S_{3,1,1} 
                +\frac{1408}{3} S_{-2,1,1} 
                +512 S_{-2,1,-2} 
                +1536 [S_{-2,2,1} + S_{-3,1,1}] 
\nonumber\\ &&               
  -3072 S_{-2,1,1,1} 
                -\frac{24 \big(
                        12+5 N+5 N^2\big) \zeta_3}{N (1+N)}
        \Biggr]
\Biggr\}
\nonumber\\ && 
+ \textcolor{blue}{C_F^2} \Biggl\{
        \textcolor{blue}{T_F N_F}  \Biggl[
                \frac{4 \big(
                        112+415 N+414 N^2+207 N^3\big)}{9 N (1+N)^2}
                +\Biggl(
                        -\frac{8 \big(
                                8+55 N+55 N^2\big)}{3 N (1+N)}
                        +\frac{1280}{9} S_2 
\nonumber\\ &&              
         -\frac{512}{3} S_3 
                        -\frac{512}{3} S_{-2,1} 
                        +128 \zeta_3
                \Biggr) S_1 
                -
                \frac{80}{3} S_2 
                -\frac{128}{3} S_2 ^2
                +\frac{1856}{9} S_3 
                -\frac{512}{3} S_4 
                +\Biggl(
                        \frac{2560}{9} S_1 
\nonumber\\ &&                 
        -\frac{256}{3} S_2 
                \Biggr) S_{-2} 
                +\Biggl(
                        \frac{1280}{9}
                        -\frac{256}{3} S_1 
                \Biggr) S_{-3} 
                -\frac{256}{3} S_{-4} 
                +\frac{256}{3} S_{3,1} 
                -\frac{2560}{9} S_{-2,1} 
                -\frac{256}{3} S_{-2,2} 
\nonumber\\ &&                
 +\frac{1024}{3} S_{-2,1,1} 
                -96 \zeta_3
        \Biggr]
        +\textcolor{blue}{C_A} \Biggl[
                \frac{P_{20}}{18 (-1+N) N (1+N)^2 (2+N)}
\nonumber\\ &&                
 +\Biggl(
                        -\frac{8 P_6}{3 (-1+N) N^2 (1+N)^2 (2+N)}
                        -\frac{4288}{9} S_2 
                        +\frac{1984}{3} S_3 
                        +320 S_4 
                        -1024 S_{3,1} 
\nonumber\\ &&              
         +\frac{1984}{3} S_{-2,1} 
                        +3712 S_{-2,2} 
                        +3840 S_{-3,1} 
                        -7168 S_{-2,1,1} 
                \Biggr) S_1
                +\Biggl(
                        256 S_3 
                        +1792 S_{-2,1} 
                \Biggr) S_1 ^2
\nonumber\\ &&                 
+\Biggl(
                        \frac{4 \big(
                                -24+151 N+151 N^2\big)}{3 N (1+N)}
                        -832 S_3 
                        -5248 S_{-2,1} 
                \Biggr) S_2 
                +\frac{352}{3} S_2 ^2
                -\frac{6160}{9} S_3 
                +\frac{2416}{3} S_4 
\nonumber\\ &&                
 +\Biggl(
                        -\frac{48 \big(
                                -5+3 N+3 N^2\big)}{(-1+N) (2+N)}
                        +\Biggl(
                                -\frac{64 P_8}{9 (-1+N) N (1+N) (2+N)}
                                -256 S_2 
                        \Biggr) S_1 
                        +
                        \frac{992}{3} S_2 
\nonumber\\ &&                
         +64 S_3 
                        +5376 S_{2,1} 
                        -384 S_{-2,1} 
                        +576 \zeta_3
                \Biggr) S_{-2} 
                +\Biggl(
                        -96
                        +512 S_1 
                \Biggr) S_{-2} ^2
\nonumber\\ &&               
 +\Biggl(
                        -\frac{32 \big(
                                243+134 N+134 N^2\big)}{9 N (1+N)}
                        +\frac{992}{3} S_1 
                        -1152 S_1 ^2
                        +2624 S_2 
                        +960 S_{-2} 
                \Biggr) S_{-3} 
                +\Biggl(
                        \frac{560}{3}
\nonumber\\ &&                    
     -1472 S_1 
                \Biggr) S_{-4} 
                +2304 S_{-5} 
                +768 S_{2,3} 
                +2688 S_{2,-3} 
                -\frac{1856}{3} S_{3,1} 
                -768 S_{4,1} 
\nonumber\\ &&                
 +\frac{64 \big(
                        243+134 N+134 N^2\big)}{9 N (1+N)} S_{-2,1}
                +\frac{992}{3} S_{-2,2} 
                -3648 S_{-2,3} 
                +1728 S_{-4,1} 
                -5376 S_{2,1,-2} 
\nonumber\\ &&                
 -1536 [S_{-2,1,-2} 
                - S_{3,1,1}] 
                - 5376 [ S_{-2,2,1} 
                + S_{-3,1,1}]  
                -\frac{3968}{3} S_{-2,1,1} 
                +10752 S_{-2,1,1,1} 
\nonumber\\ &&                
 +\frac{72 \big(
                        12+5 N+5 N^2\big) \zeta_3}{N (1+N)}
        \Biggr]
\Biggr\}
+ \textcolor{blue}{C_F^3} \Biggl\{
        \frac{P_{21}}{(-1+N) N (1+N)^2 (2+N)}
\nonumber\\ &&         
+\Biggl(
                -\frac{32 P_1}{(-1+N) N^2 (1+N)^2 (2+N)}
                +128 S_2 ^2
                -384 S_3 
                +128 S_4 
                +512 S_{3,1} 
                -384 S_{-2,1} 
\nonumber\\ &&                
 -3328 S_{-2,2}
                -3584 S_{-3,1} 
                +6144 S_{-2,1,1} 
        \Biggr) S_1 
        -256 S_{-2} ^2 S_1 
        +\Biggl(
                \frac{4 \big(
                        16+3 N+3 N^2\big)}{N (1+N)}
\nonumber\\ &&                
 +512 S_3 
                +4352 S_{-2,1} 
        \Biggr) S_2 
        -96 S_2 ^2
        +104 S_3 
        -480 S_4 
        +\Biggl(
                \frac{32 \big(
                        -5+3 N+3 N^2\big)}{(-1+N) (2+N)}
\nonumber\\ &&                
 +\Biggl(
                        \frac{128 \big(
                                -9+4 N+4 N^2\big)}{(-1+N) N (1+N) (2+N)}
                        +512 S_2 
                \Biggr) S_1 
                -192 S_2 
                +128 S_3 
                -4608 S_{2,1} 
\nonumber\\ &&                
 +256 S_{-2,1} 
                -384 \zeta_3
        \Biggr) S_{-2} 
        +\Biggl(
                \frac{576}{N (1+N)}
                -192 S_1 
                +1280 S_1 ^2
                -2176 S_2 
                -640 S_{-2} 
        \Biggr) S_{-3} 
\nonumber\\ &&         
+\Biggl(
                -96
                +1664 S_1 
        \Biggr) S_{-4} 
        -1792 S_{-5} 
        -384 S_{2,3} 
        -2304 S_{2,-3} 
        +384 S_{3,1} 
        +384 S_{4,1}
\nonumber\\ &&  
        -\frac{1152}{N (1+N)} S_{-2,1}
        -1536 S_1 ^2 S_{-2,1} 
        -192 S_{-2,2} 
        +2944 S_{-2,3} 
        -1664 S_{-4,1} 
        +4608 S_{2,1,-2} 
\nonumber\\ &&         
-768 S_{3,1,1} 
        +768 S_{-2,1,1} 
        +1024 S_{-2,1,-2} 
        +4608 S_{-2,2,1} 
        +4608 S_{-3,1,1} 
        -9216 S_{-2,1,1,1} 
\nonumber\\ &&        
-\frac{48 \big(
                12+5 N+5 N^2\big) \zeta_3}{N (1+N)}
\Biggr\}
\Biggr\}. 

\end{eqnarray}
We have calculated the transversity anomalous dimensions for the first time directly and without 
any assumptions.

The polynomials in Eqs.~(\ref{eq:ga1}--\ref{eq:ga4}) read 
\begin{eqnarray}
P_1 &=&  N^4-2 N^3-3 N^2+8 N+4,
\\
P_2 &=& 3 N^4+6 N^3-8 N^2-11 N-2,
\\
P_3 &=& 5 N^4+10 N^3+N^2-4 N-4,
\\
P_4 &=& 13  N^4+26 N^3+13 N^2-16 N-20,
\\
P_5 &=& 15 N^4+30 N^3+79 N^2+16 N-24,
\\
P_6 &=& 17 N^4+58 N^3-5 N^2-94 N-24,
\\
P_7 &=& 134 N^4+268 N^3-107 N^2-241 N+27,
\\
P_8 &=& 134 N^4+268 N^3-17 N^2-151 N-243,
\\
P_9 &=& 134 N^4+268 N^3+89 N^2-81 N-72,
\\
P_{10} &=& 134 N^4+268 N^3+116 N^2-18 N-27,
\\
P_{11} &=& 134 N^4+268 N^3+137 N^2+3 N+27,
\\
P_{12} &=& 134 N^4+268 N^3+203 N^2+69 N+27,
\\
P_{13} &=& 165 N^4+330 N^3+165 N^2+256 N+80,
\\
P_{14} &=& 245 N^4+490 N^3-117 N^2-362 N-112,
\\
P_{15} &=& 268 N^4+536 N^3+301 N^2-3 N+90,
\\
P_{16} &=& 268 N^4+536 N^3+487 N^2+183 N+126,
\\
P_{17} &=& 385 N^4+770 N^3+427 N^2+6 N-126,
\\
P_{18} &=& 389 N^4+778 N^3+398 N^2+9 N-81,
\\
P_{19} &=& 453 N^4+906 N^3+1325 N^2+344 N-264,
\\
P_{20} &=& -1359 N^5-4077 N^4-11887 N^3-14003 N^2+14494 N+13376,
\\
P_{21} &=& -29 N^5-87 N^4+227 N^3+503 N^2-230 N-256,
\\
P_{22} &=& 81 N^5+243 N^4-337 N^3-893 N^2-526 N-60,
\\
P_{23} &=& 1657 N^5+4971 N^4+4801 N^3+261 N^2-6938 N-3600,
\\
P_{24} &=& 3 N^6+9 N^5+9 N^4+9 N^3+2 N^2+4 N+2,
\\
P_{25} &=& 3 N^6+9 N^5+9 N^4+51 N^3+76 N^2+60 N+16,
\\
P_{26} &=& 22 N^6+66 N^5+95 N^4+88 N^3+197 N^2+160 N+52,
\\
P_{27} &=& 27 N^6+81 N^5-209 N^4-487 N^3-272 N^2-48 N-9,
\\
P_{28} &=& 51 N^6+153 N^5+57 N^4+35 N^3+96 N^2+16 N-24,
\\
P_{29} &=& 135 N^6-31 N^5-601 N^4-569 N^3+487 N^2+621 N+216,
\\
P_{30} &=& 209 N^6+627 N^5+627 N^4+209 N^3-108 N^2-108 N-54,
\\
P_{31} &=& 270 N^6+810 N^5-427 N^4-936 N^3+269 N^2+238 N-132,
\\
P_{32} &=& 4971 N^6+14913 N^5-24035 N^4-41453 N^3-452 N^2+7024 N-2904,
\\
P_{33} &=& -1359 N^8-5436 N^7-8274 N^6-13452 N^5-15103 N^4-12528 N^3
\nonumber\\ &&
-4120 N^2+2560 N+1584,
\\
P_{34} &=& 207 N^8+828 N^7+1491 N^6+1779 N^5+1210 N^4+453 N^3-8 N^2
\nonumber\\ &&
-160 N-72,
\\
P_{35} &=& 245 N^8+980 N^7+1542 N^6+1196 N^5+395 N^4-60 N^3+156 N^2
\nonumber\\ &&
+222 N+90,
\\
P_{36} &=&-29 N^{10}-145 N^9-226 N^8-46 N^7-N^6-469 N^5-976 N^4-940 N^3
\nonumber\\ &&
-576 N^2-208 N-32,
\\
P_{37}&=&5 N^4+10 N^3+9 N^2+4 N+4,
\\
P_{38}&=&13 N^4+26 N^3+13 N^2-16 N-20,
\\
P_{39}&=&134 N^4+268 N^3+188 N^2+54 N+45,
\\
P_{40}&=&134 N^4+268 N^3+215 N^2+45 N+54,
\\
P_{41}&=&134 N^4+268 N^3+245 N^2+111 N+135,
\\
P_{42}&=&134 N^4+268 N^3+311 N^2+177 N+135,
\\
P_{43}&=&268 N^4+536 N^3+625 N^2+321 N+414,
\\
P_{44}&=&268 N^4+536 N^3+811 N^2+507 N+450,
\\
P_{45}&=&81 N^5+162 N^4-391 N^3-286 N^2+156 N+72,
\\
P_{46}&=&3 N^6+9 N^5+9 N^4+9 N^3+6 N^2+8 N+2,
\\
P_{47}&=&3 N^6+9 N^5+9 N^4+51 N^3+12 N^2-4 N-16,
\\
P_{48}&=&22 N^6+66 N^5+95 N^4-40 N^3-115 N^2-120 N-44,
\\
P_{49}&=&27 N^6+81 N^5-155 N^4-379 N^3-92 N^2+78 N+27,
\\
P_{50}&=&135 N^6-31 N^5-481 N^4-617 N^3-395 N^2-309 N-144,
\\
P_{51}&=&165 N^6+495 N^5+495 N^4+421 N^3+144 N^2-112 N-96,
\\
P_{52}&=&209 N^6+627 N^5+627 N^4+209 N^3+36 N^2+36 N+18,
\\
P_{53}&=&453 N^6+1359 N^5+2231 N^4+1669 N^3+368 N^2+24 N+144,
\\
P_{54}&=&207 N^8+828 N^7+1443 N^6+1635 N^5+90 N^4-779 N^3-632 N^2+120,
\\
P_{55}&=&245 N^8+980 N^7+1542 N^6+1196 N^5+475 N^4+100 N^3+36 N^2+22 N-6,
\\
P_{56}&=&270 N^8+1080 N^7+347 N^6-1471 N^5-1507 N^4-417 N^3-362 N^2
\nonumber\\ &&
-12 N+144,
\\
P_{57}&=&-1359 N^{10}-6795 N^9-15246 N^8-27870 N^7-5163 N^6+40241 N^5+34648 N^4
\nonumber\\ &&
-12280 N^3-32592 N^2-17616 N-3456,
\\
P_{58}&=&-29 N^{10}-145 N^9-130 N^8+338 N^7+383 N^6+107 N^5+464 N^4+1748 N^3
\nonumber\\ &&
+1600 N^2+752 N+160,
\\
P_{59}&=&4971 N^{10}+24855 N^9+11770 N^8-70578 N^7-147665 N^6-144917 N^5
\nonumber\\ &&
-85692 N^4
-18992 N^3+22824 N^2+15840 N+2592.
\end{eqnarray}


Finally,  we turn to the non--singlet anomalous dimension 
$\gamma_{\rm NS}^{\rm (2), s}$  for which we obtain
\begin{eqnarray}
\gamma_{\rm NS}^{\rm (2), s}
&=& 4 \textcolor{blue}{N_F \frac{d_{abc} d^{abc}}{N_C}}
\Biggl[
        \frac{2 P_{60}}{(-1+N) N^5 (1+N)^5 (2+N)}
        -\frac{ P_{61}}{(-1+N) N^4 (1+N)^4 (2+N)} S_1
\nonumber\\ && 
        +\Biggl(
                -\frac{2 P_{62}}{(-1+N) N^3 (1+N)^3 (2+N)}
                -\frac{4 \big(
                        2+N+N^2\big)^2}{(-1+N) N^2 (1+N)^2 (2+N)} S_1
        \Biggr) S_{-2}
\nonumber\\ && 
        -\frac{ \big(
                2+N+N^2\big)}{N^2 (1+N)^2} 
[S_3 - 2S_{-3} + 4S_{-2,1}]
\Biggr],
\end{eqnarray}
with $d_{abc} d^{abc}/N_C = 40/9$ for $N_C = 3$ in QCD and the polynomials
\begin{eqnarray}
P_{60} &=& 
  N^8
+ 4 N^7 
+ 13 N^6 
+ 25 N^5 
+ 57 N^4 
+ 77 N^3 
+ 55 N^2 
+ 20 N 
+ 4, 
\\
P_{61} &=& 
3 N^8
+ 12 N^7 
+ 16 N^6 
+ 6 N^5 
+ 30 N^4 
+ 64 N^3 
+ 73 N^2 
+ 40 N 
+ 12, 
\\
P_{62} &=& 
N^6
+ 3 N^5 
- 8 N^4 
- 21 N^3 
- 23 N^2 
- 12 N 
- 4. 
\end{eqnarray}
Note that $\gamma_{\rm NS}^{\rm (2),s}$ has no pole at $N = 1$, but vanishes.

The splitting functions in $z$ space are given by


\noindent

For transversity the contributions to the splitting function $P_{2}^{\rm NS, +, tr}$ read 
\begin{eqnarray}
P_{\rm NS}^{(2), +, \rm tr,\delta} &=& - \Biggl\{
\textcolor{blue}{C_F^2} \Biggl[
        \textcolor{blue}{C_A} \Biggl[
                -\frac{151}{2}
                +\frac{820}{3} \zeta_2
                +\frac{1976}{15} \zeta_2^2
                +\Biggl(
                        -\frac{1688}{3}
                        -32 \zeta_2
                \Biggr) \zeta_3
                -240 \zeta_5
        \Biggr]
\nonumber\\ && 
        +\textcolor{blue}{T_F N_F} \Biggl[
                92
                -\frac{80 \zeta_2}{3}
                -\frac{928 \zeta_2^2}{15}
                +\frac{544 \zeta_3}{3}
        \Biggr]
\Biggr]
+\textcolor{blue}{C_F} \Biggl[
        \textcolor{blue}{C_A T_F N_F} \Biggl[
                -80
                +\frac{5344 \zeta_2}{27}
\nonumber\\ &&                
 -\frac{16 \zeta_2^2}{5}
                -\frac{800 \zeta_3}{9}
        \Biggr]
        +\textcolor{blue}{T_F^2 N_F^2} \Biggl[
                \frac{136}{9}
                -\frac{640 \zeta_2}{27}
                +\frac{128 \zeta_3}{9}
        \Biggr]
        +\textcolor{blue}{C_A^2} \Biggl[
                \frac{1657}{18}
                -\frac{8992 \zeta_2}{27}
\nonumber\\ &&               
 +4 \zeta_2^2
                +\frac{3104 \zeta_3}{9}
                -80 \zeta_5
        \Biggr]
\Biggr\}
+\textcolor{blue}{C_F^3} \Biggl\{
        -29
        -36 \zeta_2
        -\frac{576}{5} \zeta_2^2
        +\Biggl(
                -136
                +64 \zeta_2
        \Biggr) \zeta_3
\nonumber\\ &&     
   +480 \zeta_5
\Biggr\} 
\Biggr\} 
\delta(1-z), 
\\
P_{\rm NS}^{(2), +, \rm tr,plu} &=& - \frac{1}{1-z} \Biggl\{
        \textcolor{blue}{C_F} \Biggl[ \textcolor{blue}{T_F^2 N_F^2}
                \frac{128}{27}
                +\textcolor{blue}{C_A^2} \Biggl[
                        -\frac{4}{45} \Biggl(
                                3675
                                -2680 \zeta_2
                                +792 \zeta_2^2
                        \Biggr)
                        -\frac{176 \zeta_3}{3}
                \Biggr]
\nonumber\\ &&                 
+\textcolor{blue}{C_A T_F N_F}  \Biggl[
                        -\frac{16}{27} \Biggl(
                                -209
                                +120 \zeta_2
                        \Biggr)
                        +\frac{448 \zeta_3}{3}
                \Biggr]
        \Biggr]
        +\frac{8}{3} \textcolor{blue}{C_F^2 T_F N_F} \Biggl[
                55
                -48 \zeta_3
        \Biggr]
\Biggr\},
\nonumber\\ && 
\end{eqnarray}



In the following we use the subsidiary functions
\begin{eqnarray}
p_{qq}  &=&  \frac{1+z^2}{1+z},
\\
p_{qq}^{\rm tr}  &=&  \frac{z}{1+z}.
\end{eqnarray}
The difference terms between the $+$ and $-$-type splitting functions are given by
\begin{eqnarray}
&& 
P_{\rm NS}^{(2),+}(z) - 
P_{\rm NS}^{(2),-}(z)  =
\nonumber\\
&& 
- \Biggl\{
\textcolor{blue}{\left(C_F - \frac{C_A}{2}\right)} \Biggl\{
        \textcolor{blue}{C_F T_F N_F} \Biggl[
                -
                \frac{3904}{9} (-1+z)
                + p_{qq} \Biggl(
                        -\frac{128}{3} \HA_{-1} \HA_0^2
                        +\frac{128}{9} \HA_0^3
                        +\frac{512}{3} \HA_{-1} \HA_{0,1}
\nonumber\\ &&                
         -\frac{256}{3} \HA_{0,0,1}
                        +\frac{256}{3} \HA_{0,0,-1}
                        -\frac{512}{3} \HA_{0,1,-1}
                        -\frac{512}{3} \HA_{0,-1,1}
                        -\frac{1280}{9} \zeta_2
                        +\frac{128}{3} \HA_0 \zeta_2
                        -\frac{512}{3} \HA_{-1} \zeta_2
\nonumber\\ && 
         +128 \zeta_3
                \Biggr)
                +\frac{2624}{9} (1+z) \HA_0
                -\frac{1024 \big(
                        4+3 z+4 z^2\big)}{9 (1+z)} \HA_{-1} \HA_0
                +\frac{64 \big(
                        19+18 z+19 z^2\big) }{9 (1+z)} \HA_0^2
\nonumber\\ &&                 
+\frac{512}{3} (-1+z) \HA_1
                -\frac{256}{3} (1+z) \HA_{0,1}
                +\frac{1024 \big(
                        4+3 z+4 z^2\big)}{9 (1+z)} \HA_{0,-1}
        \Biggr]
        +\textcolor{blue}{C_F^2} \Biggl[
                \frac{7424}{9} (1-z)
\nonumber\\ && 
                + p_{qq} \Biggr(
                        \big(
                                -128 \HA_{-1}^2
                                +64 \zeta_2
                        \big) \HA_0^2
                        +\frac{320}{3} \HA_{-1} \HA_0^3
                        +\big(
                                256 \HA_{-1} \HA_0
                                +512 \HA_{-1}^2
                                +128 \zeta_2
                        \big) \HA_{0,1}
\nonumber\\ &&                
         +\big(
                                -128 \HA_0
                                -512 \HA_{-1}
                        \big) \HA_{0,0,1}
                        +512 \HA_{-1} \HA_{0,0,-1}
         +\big(
                                -256 \HA_0
                                -1024 \HA_{-1}
                        \big) \HA_{0,1,-1}
\nonumber\\ && 
                        +\big(
                                -256 \HA_0
                                -1024 \HA_{-1}
                        \big) \HA_{0,-1,1}
                        +128 \HA_{0,0,0,1}
                        +896 \HA_{0,0,0,-1}
                        +512 [\HA_{0,0,1,-1}
                        +\HA_{0,0,-1,1}
\nonumber\\ &&                        
- \HA_{0,0,-1,-1}]
                        +1024 [\HA_{0,1,-1,-1}
                        + \HA_{0,-1,1,-1}
                        + \HA_{0,-1,-1,1}]
                        +256 \HA_{-1} \HA_0 \zeta_2
                        -512 \HA_{-1}^2 \zeta_2
\nonumber\\ &&                 
        +768 \HA_{-1} \zeta_3
                \Biggr)
                +\Biggl(
                        \frac{32 \big(
                                -103-8 z+41 z^2\big)}{9 (1+z)}
                        +\frac{512 \big(
                                41+15 z+41 z^2\big)}{9 (1+z)} \HA_{-1}
\nonumber\\ &&                         
+\frac{64 \big(
                                -11-12 z+4 z^2\big)}{3 (1+z)} \zeta_2
                        -\frac{128 \big(
                                2+z^2\big)}{1+z} \zeta_3
                \Biggr) \HA_0
                +\Biggl(
                        -\frac{32 \big(
                                188+135 z+215 z^2\big)}{9 (1+z)}
\nonumber\\ &&                
         +\frac{32 \big(
                                25+24 z+25 z^2\big)}{3 (1+z)} \HA_{-1}
                \Biggr) \HA_0^2
                -\frac{160 \big(
                        2+6 z+11 z^2\big)}{9 (1+z)} \HA_0^3
                -\frac{32 z^2 }{3 (1+z)} \HA_0^4
\nonumber\\ &&                
 +\Biggl(
                        -\frac{3200}{3} (-1+z)
                        +256 (-1+z) \HA_0
                        -128 (-1+z) \zeta_2
                \Biggr) \HA_1
\nonumber\\ &&                 
+\Biggl(
                        \frac{1600 (1+z)}{3}
                        -128 (1+z) \HA_0
                        -\frac{128 \big(
                                25+24 z+25 z^2\big)}{3 (1+z)} \HA_{-1}
                \Biggr) \HA_{0,1}
\nonumber\\ &&                
 +\Biggl(
                        -\frac{512 \big(
                                41+15 z+41 z^2\big)}{9 (1+z)}
                        +704 (-1+z) \HA_0
                        +64 (-1+z) \HA_0^2
                        -\frac{128 \big(
                                3+z^2\big)}{1+z} \zeta_2
                \Biggr) \HA_{0,-1}
\nonumber\\ &&                
 +128 (1-z) \HA_{0,-1}^2
                +\frac{64 \big(
                        19+24 z+31 z^2\big)}{3 (1+z)} \HA_{0,0,1}
                +\Biggl(
                        -\frac{64 \big(
                                -41+24 z+91 z^2\big)}{3 (1+z)}
\nonumber\\ &&                       
  -\frac{128 \big(
                                3+5 z^2\big)}{1+z} \HA_0
                \Biggr) \HA_{0,0,-1}
                +\frac{128 \big(
                        25+24 z+25 z^2\big) 
                }{3 (1+z)}
[\HA_{0,1,-1} + \HA_{0,-1,1}]
\nonumber\\ &&                 
-256 (1-z) \HA_0 \HA_{0,-1,-1}
                +\frac{128 \big(
                        49-45 z+40 z^2\big)}{9 (1+z)} \zeta_2
                +\frac{128 \big(
                        25+24 z+25 z^2\big)}{3 (1+z)} \HA_{-1} \zeta_2
\nonumber\\ &&
                -\frac{64 \big(
                        7+8 z^2\big)}{1+z} \zeta_2^2
                -\frac{64 \big(
                        24+12 z+z^2\big)}{1+z} \zeta_3
        \Biggr]
\Biggr\}
+\textcolor{blue}{\left(C_F - \frac{C_A}{2}\right)^2 C_F}\Biggl\{
        \frac{11744}{9} (-1+z)
\nonumber\\ && 
        + p_{qq} \Biggl(
                -512 \HA_{-1}^2 \HA_0^2
                +128 \HA_{-1} \HA_0^3
                +\big(
                        -2048 \HA_{-1}^2
                        -512 \zeta_2
                \big) \HA_{0,1}
\nonumber\\ &&
                +2048 \HA_{-1} \HA_0 \HA_{0,-1}
                +2048 \HA_{-1} \HA_{0,0,1}
                -2048 \HA_{-1} \HA_{0,0,-1}
                +4096 \HA_{-1} [\HA_{0,1,-1}
                + \HA_{0,-1,1}]
\nonumber\\ && 
                -512 \HA_{0,0,0,1}
                +1536 \HA_{0,0,0,-1}
                -2048[ \HA_{0,0,1,-1}
                + \HA_{0,0,-1,1}
                - \HA_{0,0,-1,-1}]
                -4096 [\HA_{0,1,-1,-1}
\nonumber\\ &&                 
                + \HA_{0,-1,1,-1}
+ \HA_{0,-1,-1,1}]
                -2816 \HA_{-1} \HA_0 \zeta_2
                +2048 \HA_{-1}^2 \zeta_2
                -3072 \HA_{-1} \zeta_3
        \Biggr)
\nonumber\\ && 
        +\Biggl(
                \frac{32 \big(
                        -41+386 z+535 z^2\big)}{9 (1+z)}
                -\frac{128 \big(
                        173+78 z+173 z^2\big)}{9 (1+z)} \HA_{-1}
                +
                \frac{64 \big(
                        47+132 z+89 z^2\big)}{3 (1+z)} \zeta_2
\nonumber\\ &&                
 +\frac{128 \big(
                        7+9 z^2\big)}{1+z} \zeta_3
        \Biggr) \HA_0
        +\Biggl(
                \frac{128 \big(
                        38+29 z^2\big)}{9 (1+z)}
                +\frac{64 \big(
                        13+48 z+13 z^2\big)}{3 (1+z)} \HA_{-1}
\nonumber\\ && 
  +\frac{32 \big(
                        11+13 z^2\big)}{1+z} \zeta_2
        \Biggr) \HA_0^2
        +\frac{64 \big(
                5-18 z+8 z^2\big) \HA_0^3}{9 (1+z)}
        -\frac{8 \big(
                3+5 z^2\big) \HA_0^4}{3 (1+z)}
        +\Biggl(
                \frac{8960}{3} (-1+z)
\nonumber\\ &&             
 +512 (-1+z) \zeta_2
        \Biggr) \HA_1
        +\Biggl(
                -\frac{4480}{3} (1+z)
                +\frac{256 \big(
                        35+48 z+35 z^2\big)}{3 (1+z)} \HA_{-1}
        \Biggr) \HA_{0,1}
\nonumber\\ &&         
+\Biggl(
                \frac{128 \big(
                        173+78 z+173 z^2\big)}{9 (1+z)}
                -128 (3+13 z) \HA_0
                +\frac{64 \big(
                        1+3 z^2\big)}{1+z} \HA_0^2
                +\frac{128 \big(
                        7+5 z^2\big)}{1+z} \zeta_2
        \Biggr) 
\nonumber\\ &&  \times
\HA_{0,-1}
        +128 (-1+z) \HA_{0,-1}^2
        -\frac{128 \big(
                23+48 z+47 z^2\big)}{3 (1+z)} \HA_{0,0,1}
        +\Biggl(
                \frac{128 \big(
                        5+48 z+65 z^2\big)}{3 (1+z)}
\nonumber\\ &&                
 -\frac{128 \big(
                        7+9 z^2\big)}{1+z} \HA_0
        \Biggr) \HA_{0,0,-1}
        -\frac{256 \big(
                35+48 z+35 z^2\big)}{3 (1+z)} [\HA_{0,1,-1} + \HA_{0,-1,1}]
\nonumber\\ &&         
-\frac{256 \big(
                7+9 z^2\big)}{1+z} \HA_0 \HA_{0,-1,-1}
        +\frac{128 \big(
                23+171 z+14 z^2\big)}{9 (1+z)} \zeta_2
        -\frac{256 \big(
                35+48 z+35 z^2\big)}{3 (1+z)} \HA_{-1} \zeta_2
\nonumber\\ &&         
+\frac{32 \big(
                7+9 z^2\big)}{1+z} \zeta_2^2
        +\frac{128 \big(
                18+24 z+17 z^2\big)}{1+z} \zeta_3
\Biggr\}
\Biggr\} 
\end{eqnarray}

and
\begin{eqnarray}
&&  \hspace*{-1.5cm} P_{\rm NS}^{(2),+,\rm tr}(z) - P_{\rm NS}^{(2),-,\rm tr}(z)  = \nonumber\\ && 
-\Biggl\{\textcolor{blue}{
\left(C_F - \frac{C_A}{2}\right)} 
\Biggl\{
        \textcolor{blue}{C_F T_F N_F} \Biggr[
                \frac{448}{9} (-1+z)
                + p_{qq}^{\rm tr} \Biggl(
                        -\frac{1280}{9} \HA_0^2
                        -\frac{256}{9} \HA_0^3
                        +\Biggl(
                                \frac{5120}{9} \HA_0
\nonumber\\ &&                               
 +\frac{256}{3} \HA_0^2
                                -\frac{1024}{3} \HA_{0,1}
                                +\frac{1024}{3} \zeta_2
                        \Biggr) \HA_{-1}
                        -\frac{5120}{9} \HA_{0,-1}
                        +\frac{512}{3} [\HA_{0,0,1} - \HA_{0,0,-1}]
\nonumber\\ && 
                        +\frac{1024}{3} [\HA_{0,1,-1}
                        + \HA_{0,-1,1}]
                        +\frac{2560}{9} \zeta_2
                        -\frac{256}{3} \HA_0 \zeta_2
                        -256 \zeta_3
                \Biggl)
                -\frac{128}{3} (-1+z) \HA_1
        \Biggr]
\nonumber\\ && 
        +\textcolor{blue}{C_F^2} \Biggl[
                -\frac{4480}{9} (-1+z)
                + p_{qq}^{\rm tr} \Biggl(
                         \Biggl(
                                256 \HA_{0,0,1}
                                +1024 \HA_{0,0,-1}
                                +512 [ \HA_{0,1,-1}
                                + \HA_{0,-1,1}]
\nonumber\\ &&                
                 -\frac{64}{3} \big(
                                        -4-8 z+z^2\big) \zeta_2
                                +384 \zeta_3
                        \Biggr) \HA_0
                        -\frac{32}{9} \big(
                                -44-8 z+z^2\big) \HA_0^3
                        +\frac{32}{3} \HA_0^4
\nonumber\\ &&                
         +\Biggl(
                                \Biggl(
                                        -512 \HA_{0,1}
                                        -512 \zeta_2
                                \Biggr) \HA_0
                                -\frac{640}{3} \HA_0^3
                                +1024 [\HA_{0,0,1}
                 - \HA_{0,0,-1}]
                                +2048 [\HA_{0,1,-1}
\nonumber\\ &&                
                                + \HA_{0,-1,1}]
                                -\frac{3328}{3} \zeta_2
                                -1536 \zeta_3
                        \Biggr) \HA_{-1}
                        +\Bigl(
                                256 \HA_0^2
                                -1024 \HA_{0,1}
                                +1024 \zeta_2
                        \Biggr) \HA_{-1}^2
\nonumber\\ &&                
         +\frac{128}{3} \big(
                                -22-8 z+z^2\big) \HA_{0,0,1}
                        -256 \HA_{0,0,0,1}
                        -1792 \HA_{0,0,0,-1}
                        -1024 [\HA_{0,0,1,-1}
\nonumber\\ &&                
         + \HA_{0,0,-1,1}
                        - \HA_{0,0,-1,-1}]
                        -2048 [\HA_{0,1,-1,-1}
                        + \HA_{0,-1,1,-1} +
                        \HA_{0,-1,-1,1}]
                        -128 \HA_0^2 \zeta_2
\nonumber\\ &&                
         -256 \HA_{0,1} \zeta_2
                        +512 \HA_{0,-1} \zeta_2
                        +960 \zeta_2^2
                        +
                        \frac{64}{3} \big(
                                30-8 z+z^2\big) \zeta_3
                \Biggr)
\nonumber\\ &&                
 +\Biggl(
                        -\frac{32 \big(
                                1-14 z+3 z^2\big)}{3 (1+z)}
                        +64 (1-z) \HA_1
                        +\frac{64 \big(
                                -1+9 z-27 z^2+3 z^3\big)}{3 z} 
\nonumber\\ && 
\times \HA_{0,-1}
                \Biggr) \HA_0 
                -\frac{128 (1+z) \big(
                        1-10 z+z^2\big)}{3 z} \HA_{-1}^2 \HA_0
                -\frac{16 \big(
                        6-497 z+42 z^2+9 z^3\big)}{9 (1+z)} 
\nonumber\\ &&  \times
\HA_0^2
                +\Biggl(
                        \frac{352}{3} (-1+z)
                        -\frac{64 (-1+z) \big(
                                1-8 z+z^2\big) \zeta_2}{3 z}
                \Biggr) \HA_1
\nonumber\\ && 
                +\Biggl(
                        \frac{32 \big(
                                9+24 z-1042 z^2+24 z^3+9 z^4\big)}{9 z (1+z)} \HA_0
                        +\frac{32 \big(
                                1-8 z-44 z^2-8 z^3+z^4\big)}{3 z (1+z)} [\HA_0^2 
\nonumber\\ &&                
         - 4 \HA_{0,1}]
                        +\frac{256 (1+z) \big(
                                1-10 z+z^2\big)}{3 z} \HA_{0,-1}
                \Biggr) \HA_{-1}
                +\frac{128}{3} (1+z) \HA_{0,1}
\nonumber\\ && 
                -\frac{64 \big(
                        -1+8 z-80 z^2-56 z^3+7 z^4\big)}{3 z (1+z)} \HA_{0,0,-1}
\nonumber\\ &&       
          -\frac{32 \big(
                        9+24 z-1042 z^2+24 z^3+9 z^4\big)}{9 z (1+z)} \HA_{0,-1}
                +\frac{128 \big(
                        1-8 z-44 z^2-8 z^3+z^4\big)}{3 z (1+z)} 
\nonumber\\ && \times
[\HA_{0,1,-1} + 
\HA_{0,-1,1}]
                +
                \frac{32 \big(
                        -24-545 z+24 z^2+9 z^3\big)}{9 (1+z)} \zeta_2
\nonumber\\ && 
                -\frac{256 (1+z) \big(
                        1-10 z+z^2\big)}{3 z} \HA_{0,-1,-1}
       \Biggr]
\Biggr\}
\nonumber\\ && 
+
\textcolor{blue}{\left(C_F - \frac{C_A}{2}\right)^2 C_F}
\Biggl\{ \Biggl\{
        \frac{5728}{9} (-1+z)
        + p_{qq}^{\rm tr} \Biggl(
                \Biggl(
                        2048 \HA_{0,0,-1}
                        +4096 \HA_{0,-1,-1}
\nonumber\\ && 
                        +\frac{128}{3} \big(
                                -29-24 z+3 z^2\big) \zeta_2
                        -2048 \zeta_3
                \Biggr) \HA_0
                +\Biggl(
                        \frac{32}{9} \big(
                                -277-6 z+3 z^2\big)
\nonumber\\ &&                
         -256 \HA_{0,-1}
                        -768 \zeta_2
                \Biggr) \HA_0^2
                -\frac{1984}{9} \HA_0^3
                +\frac{64}{3} \HA_0^4
                +\Biggl(
                        \Biggl(
                                -4096 \HA_{0,-1}
                                +5632 \zeta_2
                        \Biggr) \HA_0
\nonumber\\ && 
                        +\frac{1408}{3} \HA_0^2
                        -256 \HA_0^3
                        -4096 [\HA_{0,0,1}
                        - \HA_{0,0,-1}]
                        -8192 [\HA_{0,1,-1}
                        + \HA_{0,-1,1}]
\nonumber\\ &&                  
       +6144 \zeta_3
                \Biggr) \HA_{-1}
                +\Biggl(
                        1024 \HA_0^2
                        +4096 \HA_{0,1}
                        -4096 \zeta_2
                \Biggr) \HA_{-1}^2
                -\frac{128}{3} \big(
                        -49-24 z
\nonumber\\ && 
+3 z^2\big) \HA_{0,0,1}
                +\frac{256}{3} \big(
                        -38-24 z+3 z^2\big) \HA_{0,0,-1}
                +1024 \HA_{0,0,0,1}
                -3072 \HA_{0,0,0,-1}
\nonumber\\ &&                
 +4096 [\HA_{0,0,1,-1}
                + \HA_{0,0,-1,1}
                - \HA_{0,0,-1,-1}]
                +8192 [\HA_{0,1,-1,-1}
                + \HA_{0,-1,1,-1}
\nonumber\\ &&                
 + \HA_{0,-1,-1,1}]
                +1024 \HA_{0,1} \zeta_2
                -1536 \HA_{0,-1} \zeta_2
                -512 \zeta_2^2
                +64 \big(
                        -31-8 z+z^2\big) \zeta_3
        \Biggr)
\nonumber\\ && 
        +\Biggl(
                \frac{64 \big(
                        1-22 z-5 z^2\big)}{3 (1+z)}
                +128 (9-z) z \HA_{0,-1}
        \Biggr) \HA_0
        +\frac{64 (1+z) \big(
                1-10 z+z^2\big)}{z} 
\nonumber\\ &&  \times
\HA_{-1}^2 \HA_0
        +\Biggl(
                -\frac{1088}{3} (-1+z)
                +\frac{64 (-1+z) \big(
                        1-8 z+z^2\big)}{z} \zeta_2
        \Biggr) \HA_1
\nonumber\\ &&         
+\Biggl(
                -\frac{64 \big(
                        3+12 z-518 z^2+12 z^3+3 z^4\big)}{9 z (1+z)}  \HA_0
                +\frac{128 \big(
                        3-24 z-98 z^2-24 z^3+3 z^4\big)}{3 z (1+z)} 
\nonumber\\ && 
\times
\HA_{0,1}
                -\frac{128 (1+z) \big(
                        1-10 z+z^2\big)}{z} \HA_{0,-1}
                -\frac{64 \big(
                        3-24 z-142 z^2-24 z^3+3 z^4\big)}{3 z (1+z)} \zeta_2
        \Biggr) 
\nonumber\\ &&  \times
\HA_{-1}
        -128 (1+z) \HA_{0,1}
        +\frac{64 \big(
                3+12 z-518 z^2+12 z^3+3 z^4\big)}{9 z (1+z)} \HA_{0,-1}
\nonumber\\ &&         
-\frac{128 \big(
                3-24 z-98 z^2-24 z^3+3 z^4\big)}{3 z (1+z)} [\HA_{0,1,-1} + \HA_{0,-1,1}]
\nonumber\\ &&        
 +\frac{128 (1+z) \big(
                1-10 z+z^2\big)}{z} \HA_{0,-1,-1} 
        -\frac{64 \big(
                -18-295 z-6 z^2+3 z^3\big)}{9 (1+z)} \zeta_2
\Biggr\}
\Biggr\} 
\Biggr\} 
\end{eqnarray}

in the vector and transversity cases. Here we used the shorthand notation $\HA_{\vec{a}}(z) \equiv \HA_{\vec{a}}$.
21 harmonic polylogarithms of up to weight {\sf w = 4} are contributing. 18 harmonic polylogarithms of up to weight {\sf w = 4} 
are contributing to the difference terms.

The difference terms 
$P_{\rm NS}^{(2),+}(z) - P_{\rm NS}^{(2),-}(z)$ and
$P_{\rm NS}^{(2),+, \rm tr}(z) - P_{\rm NS}^{(2),-,\rm tr}(z)$ 
do not contain soft contributions. Their expansion around $z = 1$ is given by
\begin{eqnarray}
P_{\rm NS}^{(2),+}(z) - P_{\rm NS}^{(2),-}(z)
&=&  \frac{16}{3}(1-z) \textcolor{blue}{\left(C_F- \frac{C_A}{2}\right) C_F (
\textcolor{black}{11} C_A  \textcolor{black}{ -4} T_F N_F)} + O((1-z)^2), 
\\
P_{\rm NS}^{(2),+, \rm tr}(z) - P_{\rm NS}^{(2),-,\rm tr}(z)
&=& 
-(1-z) \Biggl\{
        \textcolor{blue}{\left(C_F-\frac{C_A}{2}\right)} \Biggl[
                \textcolor{blue}{C_F^2} \Biggl(
                        -\frac{8}{9} \big(
                                -404
                                -36 \zeta_2
                        \Biggr)
                        -\frac{608}{3} \HA_1
                \Biggr)
\nonumber\\ &&
                +\frac{64}{9} \textcolor{blue}{C_F T_F N_F} \big(
                        -7+6 \HA_1\big)
        \Biggr]
        +\textcolor{blue}{\left(C_F-\frac{C_A}{2}\right)^2 C_F} \Biggl[
                -\frac{8}{9} \big(
                        116
                        +144 \zeta_2
                \big)
\nonumber\\ &&
                -\frac{8}{9} \big(
                        -696
                        +144 \zeta_2
                \big) \HA_1
        \Biggr]
\Biggr\} + O((1-z)^2). 
\end{eqnarray}
In $N$ space the leading term for $N \rightarrow \infty$ is $\propto \ln(N)/N^2$.

Finally, the splitting function $P_{\rm NS}^{(2),s}$ reads
\begin{eqnarray}
P_{\rm NS}^{(2),s}(z)  &=& - 2 \textcolor{blue}{N_F \frac{d_{abc} d_{abc}}{N_C} }
\Biggl\{
-\frac{400}{3} (1-z)
+\Biggl(
        -\frac{2}{3} (100+9 z)
        +\frac{4 (1+z) \big(
                4-7 z+4 z^2\big)}{3 z} \HA_{-1}^2
\nonumber\\ && 
        +\frac{52}{3} (1+z) \HA_{-1}
        +16 \zeta_3
\Biggr) \HA_0
+\Biggl(
        \frac{1}{3} (-38-29 z)
        -\frac{2 (1+z) \big(
                8-5 z+8 z^2\big)}{3 z} \HA_{-1}
\Biggr) \HA_0^2
\nonumber\\ && 
+\frac{2}{9} \big(
        3+8 z^2\big) \HA_0^3
-\frac{1}{3} \HA_0^4
+\Biggl(
        \frac{182}{3} (-1+z)
        +3 (1-z) \HA_0^2
\Biggr) \HA_1
+\Biggl(
        -\frac{82}{3} (1+z)
\nonumber\\ &&         
-6 (1-z) \HA_0
        +2 (1+z) \HA_0^2
        -\frac{16 (1+z) \big(
                2+z+2 z^2\big)}{3 z} \HA_{-1}
\Biggr) \HA_{0,1}
+\Biggl(
        -\frac{52}{3} (1+z)
\nonumber\\ &&         
+\frac{16 \big(
                2+3 z^2\big)}{3 z} \HA_0
        +4 (1-z) \HA_0^2
        -\frac{8 (1+z) \big(
                4-7 z+4 z^2\big)}{3 z} \HA_{-1}
\Biggr) \HA_{0,-1}
+8 (1-z) 
\nonumber\\ && 
\times
\HA_{0,-1}^2
+\Biggl(
        \frac{4}{3} \big(
                3+9 z+8 z^2\big)
        -8 (1+z) \HA_0
\Biggr) \HA_{0,0,1}
+\Biggl(
        \frac{4 \big(
                -8+3 z-21 z^2+8 z^3\big)}{3 z}
\nonumber\\ && 
        -8 (1-z) \HA_0
\Biggr) \HA_{0,0,-1}
+\frac{16 (1+z) \big(
        2+z+2 z^2\big)}{3 z} [\HA_{0,1,-1} + \HA_{0,-1,1}]
\nonumber\\ && 
+\Biggl(
        \frac{8 (1+z) \big(
                4-7 z+4 z^2\big)}{3 z}
        -16 (1-z) \HA_0
\Biggr) \HA_{0,-1,-1}
+\Biggl(
        \frac{2}{3} (67+41 z)
\nonumber\\ &&         
-\frac{2}{3} \big(
                3+27 z+32 z^2\big) \HA_0
        +2 (5+3 z) \HA_0^2
        -\frac{4 (-1+z) \big(
                4+7 z+4 z^2\big)}{3 z} \HA_1
\nonumber\\ &&         
+\frac{4 (1+z) \big(
                4-z+4 z^2\big)}{z} \HA_{-1}
        +8 (1+z) \HA_{0,1}
        -8 (1-z) \HA_{0,-1}
\Biggr) \zeta_2
-2 (3+5 z) \zeta_2^2
\nonumber\\ &&
-\frac{16}{3} \big(
        3+5 z^2\big) \zeta_3
\Biggr\}. 
\end{eqnarray}
\section{Comparison to the literature}
\label{sec:5}

\vspace*{1mm}
\noindent
We confirm the results for the non--singlet case for $\gamma_{\rm NS}^{(2),+}, \gamma_{\rm NS}^{(2),-}$  
and $\gamma_{\rm NS}^{(2),\rm s}$ in Ref.~\cite{Moch:2004pa} 
where the on--shell forward Compton amplitude has been used for the calculation. The contributions $\propto T_F$ have already 
been calculated independently as a by--product of the massive on--shell operator matrix elements 
in Ref.~\cite{Ablinger:2014vwa}. We also agree with the fixed moments, which were calculated in 
Refs.~\cite{Larin:1993vu,Larin:1996wd,Retey:2000nq,Blumlein:2004xt,Bierenbaum:2009mv} and 
the prediction of the leading $N_F$ terms for $P_{\rm NS}^{(2),+}$ and $P_{\rm NS}^{(2),-}$ computed in 
\cite{Gracey:1994nn}. 

Furthermore, we derive the small $z$ limit of the splitting functions,  given by
\begin{eqnarray}
P_{\rm NS}^{(2),+} &\simeq& 
- \Biggl\{-\frac{4}{3} \textcolor{blue}{C_F^3} \HA_0^4
+\Biggl[
        8 \textcolor{blue}{C_F^3}
        +\textcolor{blue}{C_F^2} \Biggl(
                -\frac{44 \textcolor{blue}{C_A}}{3}
                +\frac{16 \textcolor{blue}{T_F N_F}}{3}
        \Biggr)
\Biggr] \HA_0^3
+\Biggl[
        \textcolor{blue}{C_F} \Bigg[
                \frac{176 \textcolor{blue}{C_A T_F N_F}}{9}
\nonumber\\ &&
                -\frac{32}{9} \textcolor{blue}{T_F^2 N_F^2}
                +\textcolor{blue}{C_A^2} \Biggl(
                        -\frac{242}{9}
                        +60 \zeta_2
                \Biggr)
        \Biggr]
        +\textcolor{blue}{C_F^2} \Bigg[
                \frac{256 \textcolor{blue}{T_F N_F}}{9}
                +\textcolor{blue}{C_A} \Biggl(
                        -\frac{944}{9}
                        -192 \zeta_2
                \Biggr)
        \Biggr]
\nonumber\\ &&
        +\textcolor{blue}{C_F^3} \Biggl(
                -8
                +208 \zeta_2
        \Biggr)
\Biggr] \HA_0^2
+\Biggl[
        \textcolor{blue}{C_F} \Biggl(
                -\frac{704}{27} \textcolor{blue}{T_F^2 N_F^2}
                +\textcolor{blue}{C_A T_F N_F} \Biggl(
                        \frac{5072}{27}
                        -32 \zeta_2
                \Biggr)
\nonumber\\ &&
                +\textcolor{blue}{C_A^2} \Biggl(
                        -\frac{7868}{27}
                        +184 \zeta_2
                \Biggr)
        \Biggr)
        +\textcolor{blue}{C_F^2} \Biggl(
                \frac{352 \textcolor{blue}{T_F N_F}}{9}
                +\textcolor{blue}{C_A} \Biggl(
                        -\frac{740}{9}
                        -432 \zeta_2
                        -96 \zeta_3
                \Biggr)
        \Biggr)
\nonumber\\ &&
        +\textcolor{blue}{C_F^3} \big(
                60
                +384 \zeta_2
                +192 \zeta_3
        \big)
\Biggr] \HA_0
+
\textcolor{blue}{C_F^2} \Biggl[
        \textcolor{blue}{C_A} \Biggl(
                -
                \frac{1064}{9}
                -\frac{2608 \zeta_2}{9}
                +\frac{1072 \zeta_2^2}{5}
\nonumber\\ &&
                -\frac{1616 \zeta_3}{3}
        \Biggr)
        +\textcolor{blue}{T_F N_F} \Biggl(
                \frac{1300}{9}
                -\frac{640 \zeta_2}{9}
                -\frac{128 \zeta_3}{3}
        \Biggr)
\Biggr]
+\textcolor{blue}{C_F} \Biggl[
        -\frac{256}{9} \textcolor{blue}{T_F^2 N_F^2}
\nonumber\\ &&                
        +\textcolor{blue}{C_A T_F N_F} 
\Biggl(
                \frac{9896}{27}
                -32 \zeta_2
+64 \zeta_3
        \Biggr)
        +\textcolor{blue}{C_A^2} \Biggl(
                -\frac{19474}{27}
                +224 \zeta_2
                -\frac{324 \zeta_2^2}{5}
\nonumber\\ &&
                +144 \zeta_3
        \Biggr)
\Biggr]
+\textcolor{blue}{C_F^3} \Biggl(
        124
        +616 \zeta_2
        -\frac{1296 \zeta_2^2}{5}
        +384 \zeta_3
\Biggr)\Biggr\} + O(z), 
\\
  P_{\rm NS}^{(2),+} 
- P_{\rm NS}^{(2),-} 
&\simeq& 
 2 \textcolor{blue}{C_F} (\textcolor{blue}{C_A} - 2 \textcolor{blue}{C_F}) \Biggl\{
(\textcolor{blue}{C_A}
-2 \textcolor{blue}{C_F}) \HA_0^4
+\Biggl(
        -\frac{40 \textcolor{blue}{C_A}}{9}
        +\frac{32 \textcolor{blue}{T_F N_F}}{9}
\Biggr) \HA_0^3
\nonumber\\ && 
+\Biggl(
        \frac{304 \textcolor{blue}{T_F N_F}}{9}
        +8 \textcolor{blue}{C_F} \big(
                -4
                +13 \zeta_2
        \big)
        -\frac{4}{9} \textcolor{blue}{C_A} \big(
                152
                +99 \zeta_2
        \big)
\Biggr) \HA_0^2
\nonumber\\ && 
+\Biggl(
        \frac{16}{9} \textcolor{blue}{T_F N_F} \big(
                41
                +6 \zeta_2
        \big)
        +32 \textcolor{blue}{C_F} \big(
                -4
                +6 \zeta_2
                +5 \zeta_3
        \big)
        -\frac{4}{9} \textcolor{blue}{C_A} \big(
                -41
                +282 \zeta_2
\nonumber\\ &&
                +252 \zeta_3
        \big)
\Biggr) \HA_0
-8 \textcolor{blue}{C_F} \big(
        15
        -32 \zeta_2
        +7 \zeta_2^2
        -24 \zeta_3
\big)
-\frac{16}{9} \textcolor{blue}{T_F N_F} \big(
        -61
        +20 \zeta_2
\nonumber \\ &&
        -18 \zeta_3
\big)
-\frac{4}{9} \textcolor{blue}{C_A} \big(
        -367
        +92 \zeta_2
        +648 \zeta_3
        +63 \zeta_2^2
\big)
\Biggr\} + O(z), 
\\
P_{\rm NS}^{(2),s} &\simeq& 2 \textcolor{blue}{N_F \frac{d_{abc} d_{abc}}{N_C}}\Bigl[
\frac{1}{3} \HA_0^4
-\frac{2}{3} \HA_0^3
+\big(
        18
        -10 \zeta_2 \big) \HA_0^2
+\big(
        56
+        2 \zeta_2
        -16 \zeta_3
\big) \HA_0
\nonumber\\ &&
+144
-66 \zeta_2
+6 \zeta_2^2
+16 \zeta_3
\Bigr] + O(z). 
\end{eqnarray}
The leading small $z$ terms for $P_{\rm NS}^{(2),+}$ and $P_{\rm NS}^{(2),-}$ agree with the prediction in 
Ref.~\cite{Kirschner:1983di} after correcting some misprints there \cite{Blumlein:1995jp}, see also 
\cite{Bartels:1995iu}. Numerically the leading  contributions are not dominant but they are 
significantly reduced by subleading corrections, cf.~\cite{Blumlein:1995jp}. For $N_C = 3$ and $N_F = 3$ one obtains 
\begin{eqnarray}
P_{\rm NS}^{(2),+} &\simeq& 
 3.16049~\HA_0^4
+45.0370~\HA_0^3
+407.565~\HA_0^2
+1684.87~\HA_0
+3469.02, 
\\
P_{\rm NS}^{(2),-} &\simeq& 
  2.86420~\HA_0^4
+ 52.1481~\HA_0^3
+ 570.854~\HA_0^2
+ 1973.93~\HA_0
+ 3769.92. 
\end{eqnarray}
There are no predictions from genuine small $z$ calculations for subleading terms. Also the small $z$ behaviour 
of $P_{\rm NS}^{(2),s}$ has not been predicted.

The splitting functions in the case of transversity do not contain logarithmically enhanced terms in the 
small $z$ region to three--loop order, but approach the following constants
\begin{eqnarray}
 \lim_{z \rightarrow 0} P_{\rm NS}^{(2),\rm tr,+}(z) &=&  
 \frac{1}{9} \textcolor{blue}{C_F} [484 \textcolor{blue}{C_A^2} 
- 352 \textcolor{blue}{C_A T_F N_F} + 64 \textcolor{blue}{T_F^2 N_F^2}], 
\\
\lim_{z \rightarrow 0} P_{\rm NS}^{(2),\rm tr,-}(z) &=&
- \textcolor{blue}{C_F} \Biggl[
        100 \textcolor{blue}{C_A^2}
        +\frac{128}{9} \textcolor{blue}{C_A T_F N_F}
        - \frac{64}{9} \textcolor{blue}{T_F^2 N_F^2} 
\Biggr]
\nonumber\\ &&
+\textcolor{blue}{C_F^2} \Biggl[
        \textcolor{blue}{C_A} \Biggl(
                \frac{3344}{9} 
                +32 \zeta_2 \Biggr)
        - \textcolor{blue}{T_F N_F} \frac{448}{9}
\Biggl]
-\textcolor{blue}{C_F^3} 64 [
        2 + \zeta_2]. 
\end{eqnarray}
For transversity we agree with the moments $\propto \textcolor{blue}{T_F}$ calculated in \cite{Blumlein:2009rg}
and the corresponding complete $N$ and $z$--space expressions given in \cite{Ablinger:2014vwa}.
In \cite{GRACEY} the moments  1 and 3--8 of the transversity anomalous dimension have been computed, to which we
agree\footnote{In the last term of the 1st moment a factor $N_F^2$ is missing.}, as well as to the
result given in the attachment to \cite{Velizhanin:2012nm}. There the anomalous dimensions have has been obtained from 15
moments, under certain special assumptions on their mathematical structure.\footnote{There is a sign error in the term
$\propto \textcolor{blue}{C_F^2 T_F N_F}$ in Eq.~(A.15) of \cite{Velizhanin:2012nm}.} We also agree to the 16th moment of the
transversity anomalous dimension calculated in \cite{Bagaev:2012bw}.
\section{Conclusions} 
\label{sec:6} 

\vspace*{1mm}
\noindent
We have calculated the three--loop non--singlet anomalous dimensions 
$\gamma_{\rm NS}^{(2),+}$,
 $\gamma_{\rm NS}^{(2),-}$,
 $\gamma_{\rm NS}^{(2),\rm tr,+}$,
 $\gamma_{\rm NS}^{(2),\rm tr,-}$ and $\gamma_{\rm NS}^{(2),s}$ in Quantum Chromodynamics for
unpolarized and polarized deep--inelastic scattering. The method used in this first complete 
recalculation of the former results in \cite{Moch:2004pa,Velizhanin:2012nm} has been the traditional 
one, cf.~\cite{Gross:1973ju,Georgi:1951sr}, of massless off shell operator matrix elements, unlike 
the on--shell Compton amplitude at virtuality $Q^2$ in \cite{Moch:2004pa}. The present method requests 
to obtain the anomalous dimensions in a gauge--dependent framework. We confirm results given in the 
literature, also on partial results both in the unpolarized and polarized case. The former three--loop 
calculations have been performed using gauge--invariant quantities. For the non--singlet anomalous 
dimensions a finite renormalization can be avoided in the polarized case, due to a known Ward identity 
and all the results are obtained in the $\overline{\sf MS}$ scheme directly. The present calculation has 
been performed fully automatically in all its parts using a chain of dedicated codes from diagram 
generation to the final results. The three--loop anomalous dimensions have a comparatively simple 
mathematical structure, since they can be expressed in harmonic sums only \cite{Vermaseren:1998uu,Blumlein:1998if}.
We remark that also the three--loop unpolarized and polarized singlet anomalous dimensions 
$(\Delta) \gamma_{\rm PS}^{(2)}$ and $(\Delta) \gamma_{\rm qg}^{(2)}$ have been recalculated in complete form using 
the framework of massive on--shell OMEs in \cite{Ablinger:2014nga,Ablinger:2017tan,Behring:2019tus}.
The flavor non--singlet anomalous dimensions play a particular role in the associated scheme--invariant evolution 
equations for non--singlet structure functions \cite{SI}, allowing for a direct measurement of the strong coupling
constant. 

\appendix
\section{Relation between the Larin and the $\overline{\sf MS}$ scheme}
\label{sec:A}

\vspace*{1mm}
\noindent
The known Ward identity in the non--singlet case allows to derive the transformation between the Larin
scheme and the $\overline{\sf MS}$ scheme directly. We calculated the anomalous dimension $\Delta \gamma_2^{\rm 
NS,-}$
in both schemes and obtain the following transformation relations at three--loop order
\begin{eqnarray}
\Delta \gamma_2^{\rm NS,-, \overline{\sf MS}} &=& \Delta \gamma_2^{\rm NS,-, L} - 2 \beta_0\left[{z_{qq}^{(1)}}^2 
- 
2 z_{qq}^{(2),\rm NS}\right] + 2 \beta_1 z_{qq}^{(1)}, 
\\
z_{qq}^{(1)} &=& - \frac{8 \textcolor{blue}{C_F}}{N(N+1)}, 
\\
z_{qq}^{(2), \rm NS} &=& \textcolor{blue}{C_F^2} \Biggl[\frac{8 P_{63}}{N^3 (N+1)^2} + \frac{16(1+2N)}{N^2(N+1)^2} 
S_1
+ \frac{16}{N(N+1)}[S_2 + 2 S_{-2}] \Biggr]
\nonumber\\ &&
+ \textcolor{blue}{C_F C_A} \Biggl[- \frac{P_{64}}{9 N^3 (N+1)^3} - \frac{16}{N(N+1)} S_{-2}\Biggr]
+ \textcolor{blue}{C_F T_F N_F} \frac{5 N^2 - N - 3}{9 N^2 (N+1)^2}, 
\end{eqnarray}
with
\begin{eqnarray}
P_{63} &=& 2 N^4 + N^3 +8 N^2 + 5 N +2, 
\\
P_{64} &=& 103 N^4 +140 N^3 + 58 N^2 + 21 N + 36. 
\end{eqnarray}

\vspace{5mm}\noindent 
{\bf Acknowledgment.}~
We thank A.~Behring and A.~De Freitas for discussions. This project has received funding from 
the European Union's Horizon 2020 research and innovation programme under the Marie Sk\l{}odowska--Curie 
grant agreement No. 764850, SAGEX and from the Austrian Science Fund (FWF) grant SFB F50 (F5009-N15).

{\footnotesize

}
\end{document}